\documentclass[usenatbib,usegraphicx]{mn2e}
\usepackage{fleqn}
\usepackage{float}
\usepackage{color}
\usepackage{amsmath}

\def\aj{AJ}%
\def\araa{ARA\&A}%
\def\apj{ApJ}%
\def\apjl{ApJ}%
\def\aap{A\&A}%
\def\mnras{MNRAS}%
\def\nat{Nature}%
\title{Simulations of magnetic fields in isolated disk galaxies} \author[R.~Pakmor \&
V.~Springel]
{R\"udiger~Pakmor$^1$ and Volker~Springel$^{1,2}$\vspace*{0.2cm}\\
  $^1$Heidelberger Institut f\"{u}r Theoretische Studien,
  Schloss-Wolfsbrunnenweg 35, 69118 Heidelberg, Germany\\
  $^2$Zentrum f\"ur Astronomie der Universit\"at Heidelberg,
  Astronomisches Recheninstitut, M\"{o}nchhofstr. 12-14, 69120
  Heidelberg, Germany}

\begin{document}

\maketitle

\label{firstpage}

\begin{abstract}
  Magnetic fields are known to be dynamically important in the
  interstellar medium of our own Galaxy, and they are ubiquitously
  observed in diffuse gas in the halos of galaxies and galaxy
  clusters. Yet, magnetic fields have typically been neglected in
  studies of the formation of galaxies, leaving their global influence
  on galaxy formation largely unclear. Here we extend our
  magnetohydrodynamics (MHD) implementation in the moving-mesh code
  \textsc{Arepo} to cosmological problems which include radiative
  cooling and the formation of stars. In particular, we replace our
  previously employed divergence cleaning approach with a Powell
  8-wave scheme, which turns out to be significantly more stable, even
  in very dynamic environments. We verify the improved accuracy
  through simulations of the magneto-rotational instability in
  accretion disks, which reproduce the correct linear growth rate of
  the instability. Using this new MHD code, we simulate the formation
  of isolated disk galaxies similar to the Milky Way using idealized
  initial conditions with and without magnetic fields.  We find that
  the magnetic field strength is quickly amplified in the initial
  central starburst and the differential rotation of the forming disk,
  eventually reaching a saturation value. At this point, the magnetic
  field pressure in the interstellar medium becomes comparable to the
  thermal pressure, and a further efficient growth of the magnetic
  field strength is prevented.  The additional pressure component
  leads to a lower star formation rate at late times compared to
  simulations without magnetic fields, and induces changes in the
  spiral arm structures of the gas disk. In addition, we observe
  highly magnetized fountain-like outflows from the disk. These
  results are robust with numerical resolution and are largely
  independent of the initial magnetic seed field strength assumed in the
  initial conditions, as the amplification process is rapid and
  self-regulated.  Our findings suggest an important influence of
  magnetic fields on galaxy formation and evolution, cautioning
  against their neglect in theoretical models of structure formation.
\end{abstract}


\section{Introduction}

Magnetic fields are ubiquitous in the Universe and appear to be
present whenever ionized gas is involved. They are observed on vastly
different scales, from the interiors of stars to interstellar gas,
from haloes of galaxies to galaxy clusters \citep[see, e.g.,][]{Parker1979,
  Beck1996, Kulsrud1999, Carilli2002, Govoni2004}. There have even
been claims of an indirect detection of magnetic fields in cosmic
voids \citep{neronov2010a}, although the robustness of this result is
debated \citep{broderick2012a}.
     
In galaxies, magnetic fields are suspected to be particularly
important as here the magnetic pressure in the interstellar medium
(ISM) becomes comparable to the thermal pressure. Magnetic fields may
hence be dynamically relevant for the evolution of galaxies
\citep{beck2009a}, and for the regulation of their star formation. In
addition, the structure and strength of magnetic fields in galaxies
determines the propagation of cosmic rays
\citep[e.g.][]{Strong1998,Narayan2001}, which are another crucial
ingredient influencing the gas dynamics of galaxies. In fact, the
contribution of cosmic rays to the total pressure in the ISM is often
assumed to be in equipartition with the magnetic field
\citep{zweibel1997a,cox2005a}, and the coupled dynamics of both
components may be responsible for some of the galactic outflows
\citep{ipavich1975a}.
        
Magnetic field strengths have been measured for a number of galaxies
using different methods, including Zeeman splitting in maser emission
\citep{robishaw2008a} and radio polarization measurements \citep[see,
e.g.,][]{beck2007a}.  These observations provide us with quite
detailed information about the magnetic field in the Milky Way
\citep[see, e.g.,][]{jansson2012a, Jansson2012} and a few other nearby
galaxies \citep[e.g.][]{beck2007a}.
    
However, it is still not well understood how galactic magnetic fields
are originally generated and amplified, and which processes are most
relevant for their evolution \citep[for a recent review,
see][]{kulsrud2008a}. Weak initial magnetic fields could have a
cosmological origin, or are created by Biermann batteries. A further
amplification of these fields can then proceed through structure
formation flows \citep[e.g.][]{Dolag1999}, a Galactic Dynamo
\citep[e.g.][]{Hanasz2004}, or turbulent amplification
\citep[e.g.][]{Arshakian2009}.  Since the amplification of the
magnetic field in galaxies is strongly interwoven with the dynamical
state of the gas, numerical simulations offer one of the best
possibilities to study the complex magnetic field amplification and to
clarify its role in regulating star formation on the scale of whole
galaxies.
 
In clusters of galaxies, magnetic fields have been included in a
number of cosmological simulations of cluster
growth. \citet{Dolag1999, Dolag2002} have shown that the
complex shear flows and large adiabatic compression involved in
building up the intracluster gas lead to a sizable amplification of
magnetic fields, producing an end state that is largely insensitive to
the initial seed field strength and configuration, and qualitatively
matches observed Faraday rotation maps.

So far, only few high-resolution simulations of galaxy formation have
attempted to include magnetic fields, largely owing to the technical
challenges involved. \citet{wang2009b} studied the evolution of
magnetic fields in the formation of an isolated dwarf galaxy with
radiative cooling, but without star formation.  \citet{dubois2010a}
studied magnetic fields in dwarf galaxies including cooling and star
formation, highlighting the dispersal of magnetic fields into the
intergalactic medium (IGM) by supernova driven wind.  These groups
used a finite volume discretization of magnetohydrodynamics (MHD), but
there have also been attempts to employ particle-based representations
of MHD. In particular, \citet{Kotarba2009} compared two different MHD
smoothed particle hydrodynamics (SPH) codes in simulations of magnetic
field amplification in spiral galaxies.  \citet{Kotarba2010,
  Kotarba2011} applied MHD-SPH techniques in studies of the magnetic
field amplification in galaxy mergers, and \citet{beck2012a} employed
SPH to simulate the evolution of the magnetic field in a Milky
Way-like halo embedded in a cosmological environment. However, the
latter study focussed on predictions for the diffuse gas in the
galactic halo and did not discuss the properties of the magnetic field
on the scale of the gaseous disk.
    
In this paper, we aim to investigate the role of magnetic fields in
the evolution of Milky Way-sized galaxies, including the regime where
magnetic forces become dynamically important in the dense ISM gas. We
shall focus on idealized isolated galaxy models in this work, in
preparation for future cosmological simulations of galaxy formation
that include magnetic fields. Another goal of the present study lies
in describing and demonstrating the technical improvements in our
updated MHD solver, which now reaches an accuracy that allows it to
solve difficult MHD hydrodynamical problems such as field
amplification through the magnetorotational instability.

Accounting for ideal magnetohydrodynamics in the context of galaxy
formation simulations is technically challenging, primarily because of
the well-known difficulties to maintain the $\nabla\cdot \textbf{B}=0$
constraint in simple discretization schemes, an issue which is
especially severe in the light of the very large dynamic range and
spatial adaptively that is required in galaxy simulations.  One of the
best approaches to address this problem lies in constrained transport
schemes coupled to adaptive mesh refinement (AMR). However, this
technique also comes with disadvantages such as comparatively large
advection errors for supersonic bulk flows, something quite common in
cosmic structure formation. The moving-mesh code \textsc{Arepo}
\citep{springel2010a} is a new approach that overcomes this
limitation, but requires a more complicated unstructured mesh
geometry.  Recently, we presented a first implementation of MHD in the
\textsc{Arepo} code \citep{pakmor2011d}, which combines the accuracy of
mesh-based techniques for hydrodynamics with the automatic adaptivity
and geometric flexibility of particle based techniques such as
SPH. Here, we shall first begin by presenting an important improvement
of our previous implementation of ideal MHD, as well as describing the
required extensions for cosmological integration in comoving
coordinates. We will also show validation tests of the technique that
demonstrate that the $\nabla\cdot \textbf{B}$ errors are kept under
control and that complicated flows such as strong field amplification
in shear flows are calculated correctly. This method is then applied
to simulations of isolated galaxy models, with and without magnetic
fields.

This paper is structured as follows.  In
Section~\ref{sec:implementation}, we describe the technical
implementation of our improved MHD solver. In Section~\ref{sec:mri} we
apply our code to simulate a magnetized accretion disk and compare the
linear growth of the magneto-rotational instability to the analytical
solution and the results of previous work in the literature. In
Section~\ref{sec:setup}, we briefly summarize the implementation of
cooling, star formation, and associated feedback in our runs, and we
specify our initial conditions.  Section~\ref{sec:results} is devoted
to a discussion of our primary results for high-resolution simulations
of isolated Milky Way-sized dark matter halos with and without
magnetic fields. Finally, we give a summary and discussion of our
findings in Section~\ref{sec:summary}.

\section{MHD Implementation}

\label{sec:implementation}

\subsection{Cosmological ideal magnetohydrodynamics}

The equations of ideal MHD in physical coordinates $\textbf{r}$ are
given by
\begin{eqnarray}
  \label{eqn:mhd}
  & & \frac{\partial \rho}{\partial t} + \nabla_{\rm r} \cdot \left( \rho \textbf{v} \right) = 0 ,\\
  & & \frac{\partial \rho \textbf{v}}{\partial t} +  \nabla_{\rm r} \cdot \left( \rho \textbf{v} \textbf{v}^T + p_{\rm tot} - \textbf{B} \textbf{B}^T \right) = 0, \\
  & & \frac{\partial E}{\partial t} + \nabla_{\rm r} \cdot \left[  \textbf{v}\left(E  + p_{\rm tot} \right) - \textbf{B} \left( \textbf{v} \cdot \textbf{B} \right) \right] = 0 ,\\
  & & \frac{\partial \textbf{B}}{\partial t} +  \nabla_{\rm r} \cdot \left( \textbf{B} \textbf{v}^T - \textbf{v} \textbf{B}^T \right) = 0.
\end{eqnarray}
The time derivatives are here defined at constant physical position
$\textbf{r}$, and the spatial derivatives in $\nabla_{\rm r}$ are
defined with respect to $\textbf{r}$.  Here, $p_{\rm tot} = p_{\rm
  gas} + \frac{1}{2} \textbf{B}^2$ is the total gas pressure, and $E =
\rho u_{\rm th} + \frac{1}{2} \rho \textbf{v}^2 + \frac{1}{2}
\textbf{B}^2$ is the total energy per unit volume, with $u_{\rm th}$
denoting the thermal energy per unit mass. $\rho$, $\textbf{v}$ and
$\textbf{B}$ represent the local gas density, velocity and magnetic
field strength, respectively.
    
In addition, the magnetic field $\textbf{B}$ has to fulfill the
divergence constraint, $\nabla\cdot \textbf{B}= 0$. Analytically, this
constraint will automatically be met at all times if it is fulfilled
by the initial magnetic field. Discretization errors, however, can
lead to a non-vanishing divergence of the magnetic field in numerical
simulations. We will discuss this issue and how we deal with it in
Section~\ref{sec:divb}.
    
\begin{figure*}
  \centering
  \includegraphics[width=19cm,trim=30 10 0 20,clip]{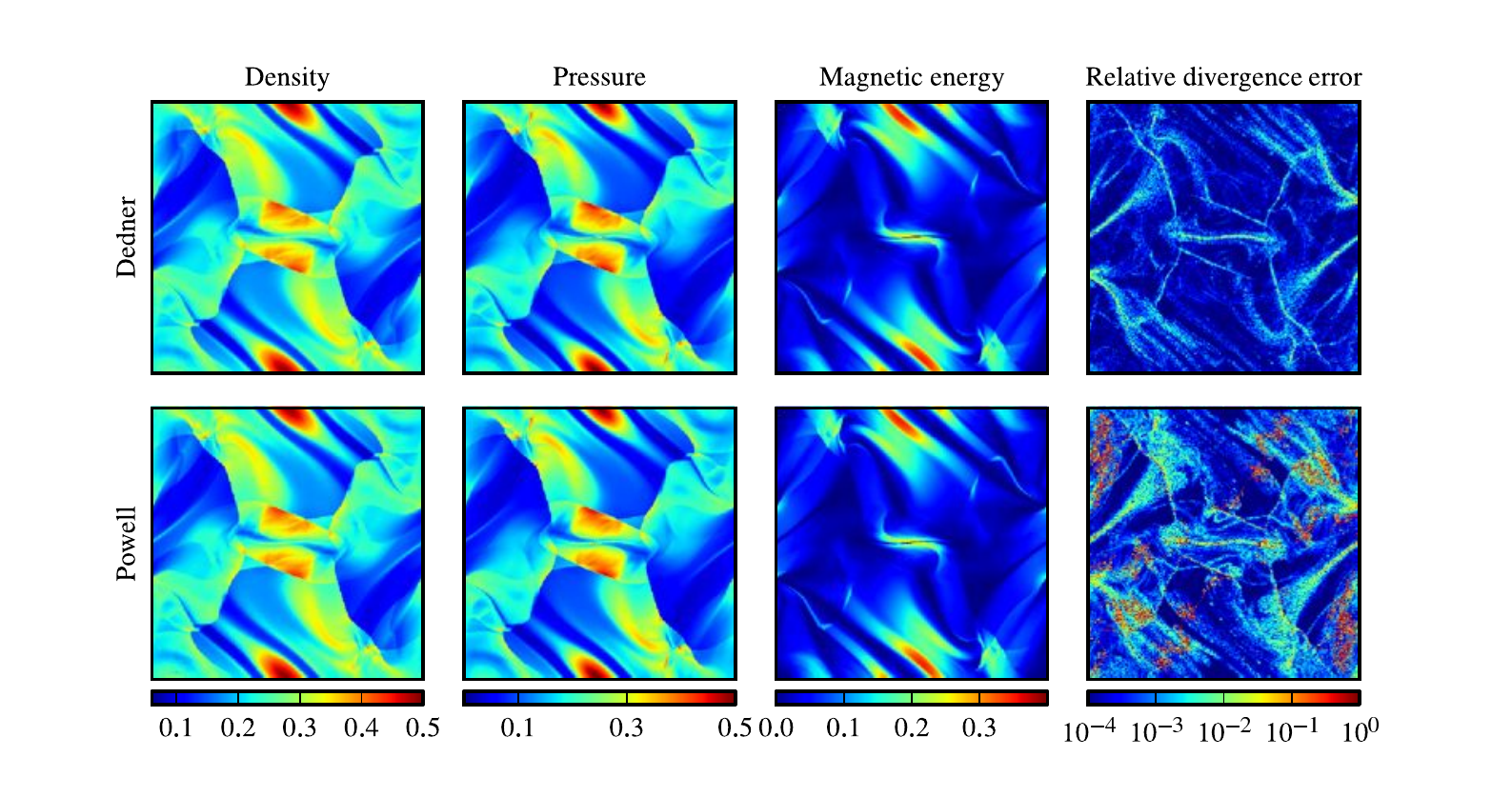}
  \caption{Orszag-Tank vortex at $t=0.5$ for the Dedner divergence
    cleaning (top row) and the Powell scheme (bottom
    row). Shown from left to right are density, pressure, magnetic
    energy, and the absolute value of the relative divergence
    error as defined in Eqn.~(\ref{eqn:divberror}). Both simulations
    were run with $600 \times 600$ cells.}
  \label{fig:divbtest}
\end{figure*}

\subsection{Extending the MHD equations to comoving coordinates}

For definiteness, we briefly discuss the formulation of the MHD
equations adopted in our code for expanding space.  In cosmological
simulations, it is convenient and common practice to use spatial
coordinates comoving with the expansion of the Universe, such that
only gas motions relative to the local standard of rest need to be
computed. As usual, we parametrize the global expansion of the
Universe with the time-dependent scale-factor $a(t)$. Besides
introducing comoving coordinates $\textbf{x}$, we also define other
`comoving' variables (denoted with a `c' index) for velocity, density,
pressure, and magnetic field, as follows:
\begin{eqnarray}
      && \textbf{r} \; = \; a\,\textbf{x} ,\\
      && \textbf{u} \; = \; \textbf{v} -   \dot{a}\, \textbf{x} ,\\
     && \rho \; = \; \rho_{\rm c}\, a^{-3}, \\
     && p \; = \; p_{\rm c}\, a^{-3}, \\
     && \textbf{B} \; = \; \textbf{B}_{\rm c}\, a^{-2}.
\end{eqnarray}
Here $\textbf{v} = \dot{\textbf{r}}$ is the physical velocity, and
$\textbf{u}$ the peculiar velocity.  With the exception of the
magnetic field, these replacements are the standard definitions, but different
choices have been used in the literature to define
$\textbf{B}_{\rm c}$. The definition we adopt here has the advantage
of avoiding a source term in the induction equation, as shown below.

The transformed equations in comoving coordinates based on these 
replacements are as follows:
\begin{eqnarray}
  &&\frac{\partial \rho_{\rm c}}{\partial t} + \frac{1}{a} \nabla_{\rm
    x} \cdot \left( \rho_{\rm c} \textbf{u} \right) = 0, \nonumber \\
  &&\frac{\partial \rho_{\rm c} \textbf{u}}{\partial t} + \frac{1}{a} \nabla_{\rm x} \cdot \left( \rho_{\rm c} \textbf{u} \textbf{u}^T +
    p_{\rm tot, c} - \frac{\textbf{B}_{\rm c} \textbf{B}_{\rm c}^T}{a}
  \right) = \nonumber \\
  && = - \frac{\dot{a}}{a} \rho_{\rm c} \textbf{u} , \\
  &&\frac{\partial E_{\rm c}}{\partial t} + \frac{1}{a} \nabla_{\rm x} \cdot \left[  \textbf{u}\left(E_{\rm c}  +
      p_{\rm tot, c} \right) - \textbf{B}_{\rm c} \left( \textbf{u}
      \cdot \textbf{B}_{\rm c} \right) \right] =  \nonumber \\
  && =  - \frac{\dot{a}}{a} \left( \rho_{\rm c} \textbf{u}^2 + 2 u_{\rm th} + \frac{\textbf{B}^2_{\rm c}}{2a} \right), \\
  &&\frac{\partial \textbf{B}_{\rm c}}{\partial t} + \frac{1}{a} \nabla_{\rm x} \cdot
  \left( \textbf{B}_{\rm c}  \textbf{u}^T - \textbf{u} \textbf{B}_{\rm c}^T \right) = 0.
\end{eqnarray}
Now the time derivatives are defined at constant comoving position
$\textbf{x}$ and the spatial derivatives in $\nabla_{\rm x}$ are
defined with respect to $\textbf{x}$. We also introduced the total
comoving pressure $p_{\rm tot, c} = p_{\rm gas} + \frac{1}{2a}
\textbf{B}^2_{\rm c}$ and the total comoving energy density per unit
volume, $E_{\rm c} = \rho_{\rm c} u_{\rm th} + \frac{1}{2} \rho_{\rm
  c} \textbf{u}^2 + \frac{1}{2a} \textbf{B}_{\rm c}^2$.

To get rid of most of the source terms we finally introduce scaled
variables for the momentum and energy, viz.
\begin{eqnarray}
  && \textbf{w} = a \textbf{u}, \\ 
  && {\cal E} = a^2 E_{\rm c}.
\end{eqnarray}    
Substituting these new variables into the cosmological
MHD equations leads to the conservative system
we actually solve in our code:
\begin{eqnarray}
  \label{eqn:mhdfinal}
 && \hspace*{-0.5cm} \frac{\partial \rho_{\rm c}}{\partial t} + \frac{1}{a} \nabla_{\rm
    x} \cdot \left( \rho_{\rm c} \textbf{u} \right) = 0,  \\
&& \hspace*{-0.5cm} \frac{\partial \rho_{\rm c} \textbf{w}}{\partial t} + \nabla_{\rm x} \cdot \left( \rho_{\rm c} \textbf{u} \textbf{u}^T +
    p_{\rm tot, c} - \frac{\textbf{B}_{\rm c} \textbf{B}_{\rm c}^T}{a}
  \right) = 0,  \\
&& \hspace*{-0.5cm} \frac{\partial {\cal E}}{\partial t} + a \nabla_{\rm x} \cdot \left[  \textbf{u}\left(E_{\rm c}  +
      p_{\rm tot, c} \right) - \frac{1}{a} \textbf{B}_{\rm c} \left( \textbf{u} \cdot \textbf{B}_{\rm c} \right) \right] = 
    \frac{\dot{a}}{2} \textbf{B}^2_{\rm c},  \\
&& \hspace*{-0.5cm} \frac{\partial \textbf{B}_{\rm c}}{\partial t} + \frac{1}{a} \nabla_{\rm x} \cdot
  \left( \textbf{B}_{\rm c}  \textbf{u}^T - \textbf{u} \textbf{B}_{\rm
      c}^T \right) = 0. 
  \label{eqn:induction}
\end{eqnarray}
Note that in these equations only one source term remains as part of
the energy equation.
 
The above equations have a similar form as the ordinary MHD equations
(\ref{eqn:mhd}) in fixed coordinates, and in fact reduce to them for
$a(t) = 1$, as expected. Note that we can use a normal Riemann solver
to calculate the fluxes without the need for cosmological adjustments,
using the following procedure:
\begin{enumerate}
\item We reconstruct the comoving primitive variables for a 
  cell-interface in its rest-frame, using the MUSCL-Hancock approach as
  described in \citet{springel2010a}.
\item We scale the magnetic field as $\textbf{B}^\prime_{\rm c} =
  {\textbf{B}_{\rm c}}/{\sqrt{a}}$.
\item We then calculate the fluxes across the moving interface from the
  reconstructed primitive variables and $\textbf{B}^\prime_{\rm c}$
  using the HLLD riemann solver \citep{miyoshi2005a}, as described in
  \citet{pakmor2011d}.
\item We finally revert the scaling of the magnetic field
fluxes using
  $\textbf{F}_{B} = \textbf{F}^\prime_{B} \times \sqrt{a}$, and
  multiply all other fluxes with the appropriate powers of the scale
  factor $a$.
\end{enumerate}
The source term in the energy equation (\ref{eqn:mhdfinal}) is treated
in a Strang-split fashion by applying two half-timesteps before and
after evolving the homogenous system by one step, similar as the
treatment of the gravitational source terms.

\subsection{The divergence constraint} \label{sec:divb}

Although the induction equation (\ref{eqn:induction}) analytically
fulfills the divergence constraint of an initially divergence free
magnetic field at all times, this property is usually lost
for the discretized version of the induction equation. However, a
magnetic field which develops a non-zero divergence can lead to severe
artifacts in numerical simulations. In particular, divergence errors
can induce a huge local amplification of the magnetic and velocity
fields, rendering both unphysical.

An elegant solution to this problem is the constraint transport
approach \citep{evans1988a}. This scheme discretizes magnetic and
electric fields such that the sequence of their update steps
manifestly guarantees a vanishing divergence of the magnetic field at
cell centres, to machine precision. Unfortunately, however, this
method is at present limited to Cartesian grids, and it is conceptually
unclear whether it can be adapted to unstructured moving meshes at
all.

Instead of trying to completely avoid divergence errors by
construction, as in constraint transport, we therefore attempt to keep
them as small as possible, so that they do not affect our results. In
\citet{pakmor2011d} we implemented the Dedner divergence cleaning
approach \citep{dedner2002a} in the moving-mesh code {\small AREPO}
and showed that it is usually able to keep the divergence error at a
small level. In the Dedner scheme, a local divergence error is both
advected away from the place where it originates and also damped at
the same time.  However, as shown in \citet{pakmor2011d}, a relatively
restrictive timestep criterion is required to use the Dedner scheme
together with individual timesteps for the cells, and even then it is
difficult to guarantee stability in very dynamic environments,
encumbering applications to cosmic structure formation.
    
To remedy these issues, we now adopt the Powell approach for divergence
control \citep{powell1999a}.  In this scheme, additional source
terms are introduced into the momentum equation, induction equation
and energy equation:
\begin{eqnarray}
      &&\frac{\partial \rho_{\rm c}}{\partial t} + \frac{1}{a} \nabla_{\rm x} \cdot \left( \rho_{\rm c} \textbf{u} \right) = 0 ,\\
      &&\frac{\partial \rho_{\rm c} \textbf{w}}{\partial t} + \nabla_{\rm x} \cdot \left( \rho_{\rm c} \textbf{u} \textbf{u}^T +
                p_{\rm tot, c} - \frac{\textbf{B}_{\rm c}
                  \textbf{B}_{\rm c}^T}{a} \right) = \nonumber \\
      &&   \hspace*{1cm}  - \frac{1}{a} \left( \nabla_{\rm x} \cdot
        \textbf{B}_{\rm c} \right) \textbf{B}_{\rm c} , \\
      &&\frac{\partial {\cal E}}{\partial t} + a \nabla_{\rm x} \cdot \left[  \textbf{u}\left(E_{\rm c}  +
              p_{\rm tot, c} \right) - \frac{1}{a} \textbf{B}_{\rm c}
            \left( \textbf{u} \cdot \textbf{B}_{\rm c} \right) \right]
          = \nonumber \\
      &&  \hspace*{1cm} \frac{\dot{a}}{2} \textbf{B}^2_{\rm c} 
            - \frac{1}{a} \left( \nabla_{\rm x} \cdot \textbf{B}_{\rm c} \right) \left( \textbf{u} \cdot \textbf{B}_{\rm c} \right), \\
      &&\frac{\partial \textbf{B}_{\rm c}}{\partial t} + \frac{1}{a} \nabla_{\rm x} \cdot
              \left( \textbf{B}_{\rm c}  \textbf{u}^T - \textbf{u} \textbf{B}_{\rm c}^T \right) = 
              - \frac{1}{a} \left( \nabla_{\rm x} \cdot \textbf{B}_{\rm c} \right) \textbf{u}.
\end{eqnarray}
These source terms encode a passive advection of $\nabla \cdot
\textbf{B} / \rho$ with the flow and counteract further growth of
local $\nabla\cdot \textbf{B}$ errors. By experience, we find that this
scheme is very stable in practical applications. A welcome advantage
is also that it is completely local and does not require any
additional constraints on the timesteps, even when all cells are
evolved on individual timesteps.

\begin{figure}
  \centering
  \includegraphics[trim=20 10 20 2, width=8.0cm]{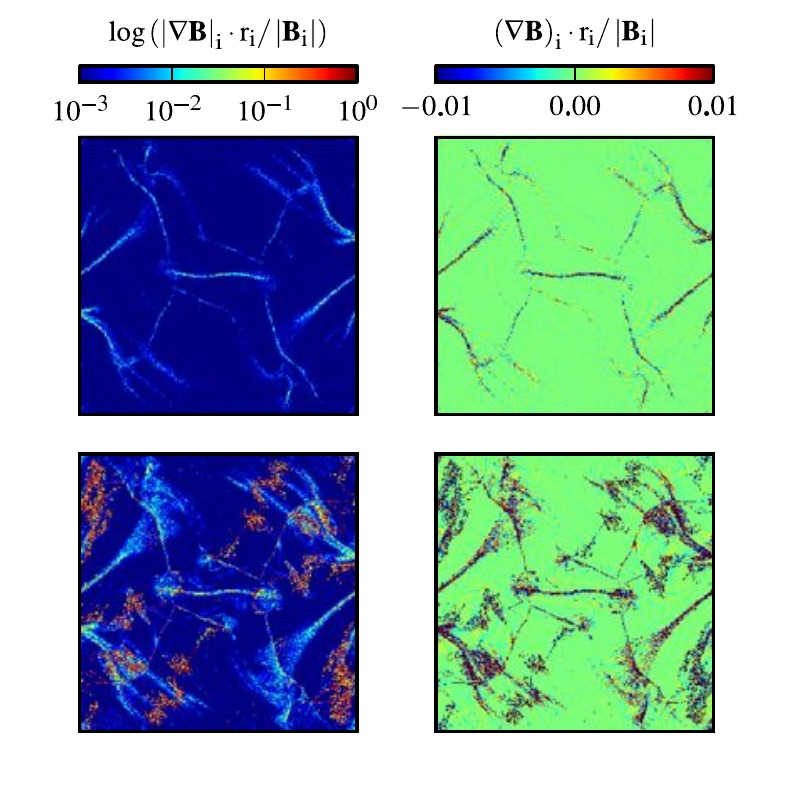}
  \caption{Maps of the divergence error for a Orszag-Tank vortex at
    $t=0.5$ for the Dedner divergence cleaning (top row) and the
    Powell scheme (bottom row). The left column shows the absolute
    value of the relative magnetic divergence error on a logarithmic
    scale, whereas the right column shows its signed value on a linear
    scale.  Both simulations were run with $600 \times 600$ cells.}
      \label{fig:divbtest2}
\end{figure}

We apply the divergence source terms for all active cells at the same
time when we calculate and apply the local fluxes. In each timestep,
we first define the magnetic field perpendicular to a cell interface as
the average of the perpendicular magnetic fields left and right of the
interface, $B_x = \frac{1}{2} \left( B_{x,L} + B_{x,R} \right)$. This
value of $B_x$ is then used in the Riemann solver. We estimate the
divergence of the magnetic field of an active Voronoi cell by
\begin{equation}
  \label{eqn:divberror}
  \nabla \cdot \textbf{B}_i = \frac{1}{V_i} \sum_\mathrm{faces} \textbf{B} \cdot \hat{\textbf{n}} \, A_i,
\end{equation}
using the average normal component $B_x$ at the interfaces calculated
as above, and add the fluxes and divergence source terms at the same
time to our conservative variables. To estimate the relative size of
the divergence error, we typically also use the values for the
divergence of the magnetic field calculated as part of this treatment
of the source terms.

\subsection{Testing the divergence control}

As one important test of our new implementation of divergence control
based on the Powell approach, we simulate the Orszag-Tang vortex
problem with the same configuration as shown in \citet{pakmor2011d},
using both the previously implemented Dedner divergence cleaning and
the Powell approach. The two simulations are shown at time $t=0.5$ in
Fig.~\ref{fig:divbtest}, for comparison. There is excellent agreement
between both results in density, pressure and magnetic energy,
although the employed approach to limit divergence errors is
significantly different. Note that this agreement is a non-trivial
outcome, given that small and seemingly innocent changes in the
implementation details of each of the divergence control methods are
typically readily punished by clearly visible artifacts in these
physical quantities. This is because both approaches are intricate
numerical schemes that just barely keep the $\nabla\cdot \textbf{B}$
related errors under control.
      
There are, however, some differences in the relative size of the
resulting divergence errors. As is clearly visible in
Fig.~\ref{fig:divbtest}, the absolute value of the relative divergence
error is larger in some regions for the Powell approach compared to
the Dedner divergence cleaning. This is visible even better in
Fig.~\ref{fig:divbtest2}, where maps of the divergence error are
shown. At time $t=0.5$, we find a volume-weighted average of the
magnitude of the relative divergence error equal to $9 \times 10^{-4}$
for the Dedner divergence cleaning, and $2.4 \times 10^{-2}$ for the
Powell approach, respectively. If the sign of the divergence error is
not ignored in the averaging, the mean divergence error is much
smaller, only $-3 \times 10^{-5}$ for the Dedner scheme and $-2 \times
10^{-4}$ for the Powell method, because positive and negative
fluctuations in $\nabla\cdot \textbf{B}$ tend to largely cancel.

It is important to note that for both measurement methods the average
relative divergence error is much smaller than unity and spatially
highly localized. The spatial distribution of the errors is very
similar for both divergence cleaning schemes. Interestingly, in places
where the absolute value of the relative divergence error becomes
large, its sign alternates between neighboring cells.  This suggests
that the divergence error occurs as a locally confined problem, most
likely originating in approximate solutions of the Riemann problems
and in the local discretization and truncation errors. The local
errors are oscillatory in nature, and the partial cancellation among
neighboring cells helps to keep any large-scale impact very small.
Although the average divergence error is about an order of magnitude
larger for the Powell scheme compared to the Dedner cleaning, there
are no noticeable differences in any fluid quantities, as shown in
Fig.~\ref{fig:divbtest}, supporting this interpretation.  This good
alignment of the two schemes in the Orszag-Tang vortex problem is only
lost at much later time, when the initial vortex has decayed into
turbulence, and the turbulent state as a function of time becomes
slightly different between the schemes. In Sec.~\ref{sec:divbgal}, we
will return to a discussion of the divergence error in our
applications of the code to realistic galaxy simulation.

\begin{figure}
  \centering
  \includegraphics[trim=15 5 12 1, width=8cm]{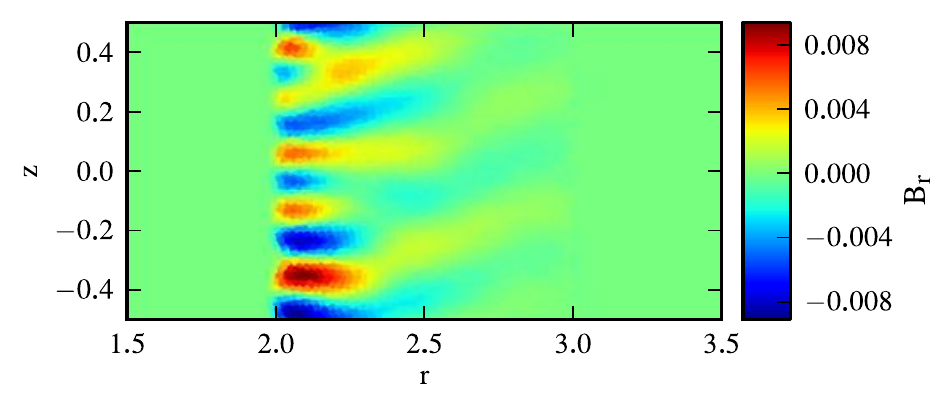}
  \caption{Radial magnetic field at $t=30$ for a test of the
    magneto-rotational instability, after about five inner orbits.}
  \label{fig:mri_br}
\end{figure}

\section{Magneto-rotational instability}  \label{sec:mri}

The magneto-rotational instability (MRI) \citep{balbus1991a} is one of
the most important MHD phenomena in astrophysics, making it a critical
test-problem for numerical MHD codes. In particular, in accretion
disks the MRI is suspected to provide the primary means for an
efficient transport of angular momentum in the radial direction,
allowing the gas to be accreted by the central compact object. The MRI
may also facilitate the growth of magnetic fields in galaxies
\citep{Kitchatinov2004}.  Interestingly, for accretion disks it is
possible to calculate the linear growth rate of the MRI analytically
\citep{balbus1991a}, allowing for a direct quantitative check of
whether a numerical MHD implementation can correctly account for this
process.
       
\subsection{Test setup}

We employ the same setup as used in \citet{flock2010a}, but simulate
the full disk to avoid the need for azimuthal boundary conditions.  In
this setup, a thin, non self-gravitating, Keplerian disk rotates
around a central body of unit mass, $M=1$, implying a centripetal
acceleration of $g(R) = 1 / R^2$ in cylindrical coordinates. The disk
extends from $R=1$ to $R=4$ in radial direction and from $z=-0.5$ to
$z=0.5$ in vertical direction. Initially, it has uniform density $\rho
=1$ and pressure $P=c_s^2\rho / \gamma$, with $c_s = 0.1$ and $\gamma
= 5/3$, and rotates with an azimuthal velocity of $v_\phi = \sqrt{ 1 /
  R }$. The radial and vertical velocity is given by axisymmetric
random perturbations of amplitude $\pm 5 \times 10^{-4}$.
       
A uniform vertical magnetic field with a strength of $0.05513 / n$ and
$n = 4$ pervades the disk between the radii $R=2$ and 
$R=3$. We use periodic boundary conditions in the vertical direction
and a layer of boundary cells at $R=1$ and $R=4$. These boundary cells
keep the undisturbed parts of the disk at their initial radial
position.

\begin{figure}
  \centering
  \includegraphics[trim=20 5 20 5, width=8cm]{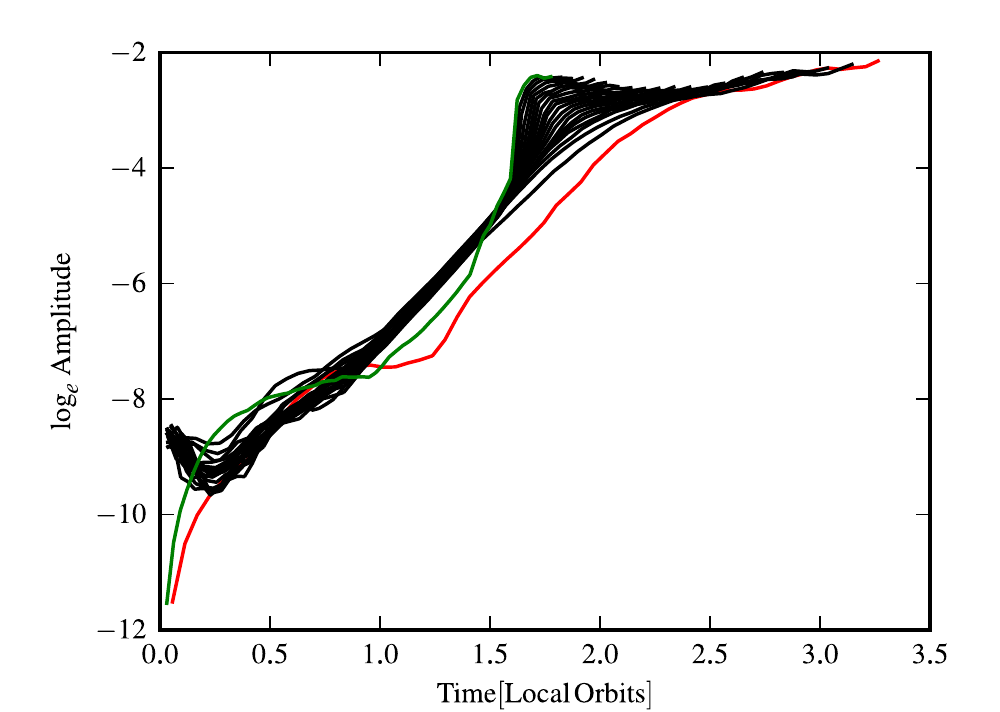}
  \caption{Maximal amplitude of the radial magnetic field at $21$
    equally spaced radii from $R = 2$ to $R = 3$ versus local
    orbits. The red and green line show the amplitude at radii $R = 2$
    and $R = 3$, respectively.}
  \label{fig:mri_brmax}
\end{figure}
       
The mesh-generating points are initially arranged on $128$ circular
rings in the plane of rotation and duplicated $64$ times uniformly in
the vertical direction. The number of cells per ring is set such that
cells have a mass of $\simeq 10^{-5}$ per cell, with a
uniform spacing within each ring.
       
For this setup, the radial and azimuthal magnetic field should be
amplified exponentially in time according to $B = B_0\, \exp
(\gamma_{\rm{MRI}}\, t)$ \citep{balbus1991a}.  The fastest-growing
mode in vertical direction should be the $n=4$ mode, the dominant
radial mode depends on the initial perturbations. The critical
(fastest possible) mode grows with $\gamma_{\rm{MRI}} = 0.75\,
\Omega$, with an angular frequency $\Omega = R^{-1.5}$ as a function
of radius.

\subsection{Results}
   
Figure~\ref{fig:mri_br} shows the radial magnetic field after about
$5$ inner orbits, corresponding to $1.8$ orbits at $R=2$. The MRI is
clearly active and a radial magnetic field has emerged. The dominant
vertical mode of the instability is close to $n=4$, as predicted for
this setup. The magnetic field is largest at $R=2$ and becomes smaller
with increasing radius \citep[for the pattern, compare
to][]{flock2010a}. We note that the "checkerboard"-instability that
occurs for some implementations tested in \citet{flock2010a} does not
show up in our simulation.
       
The growth of the amplitude of the radial magnetic field at different
radii is shown in Fig.~\ref{fig:mri_brmax}. Between about $0.5$ and
$1.5$ local orbits, the amplitude increases exponentially, essentially
at all radii. Only the very inner and outer radii at $R=2$ and $R=3$
are slightly different, most likely due to boundary effects. Starting
after about $1.5$ local orbits, the growth becomes steeper at a
progressively larger range of radii, since the magnetic field is
already saturated in the inner parts of the disk, which leads to
non-linear growth in regions close to the saturated part.

\begin{figure}
  \centering
  \includegraphics[trim=20 5 20 5, width=8cm]{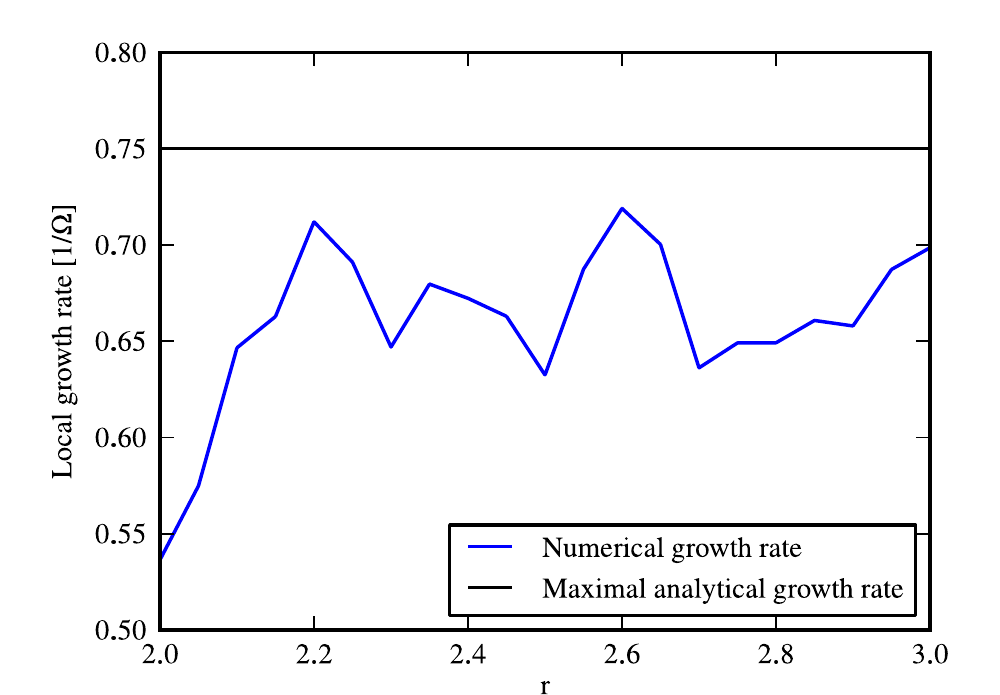}
  \caption{Exponential growth rate of the magnetic field amplitude
    measured at different radii (blue line). The black line shows the
    maximum growth rate expected analytically.}
  \label{fig:mri_growth}
\end{figure}

We fit the growth rate at different radii selected between $1.0$ and $1.5$
local orbits and show the result in Fig.~\ref{fig:mri_growth}.  The
growth rate in units of the inverse orbital period ranges from $0.53$
to $0.72$ at different radii with an average growth rate of $0.66$, in
very good agreement with the results by \citet{flock2010a} for the
same setup, which were based on completely different MHD
implementations that included constraint transport. Also, there are no
regions in our results where the local growth rate exceeds the fastest
possible growth rate of $0.75$. Reassuringly, we thus conclude that
our code is able to correctly reproduce the linear phase of the
magneto-rotational instability.
       
\begin{figure*}
  \centering
  \includegraphics[width=18cm,trim=40 50 5 30,clip]{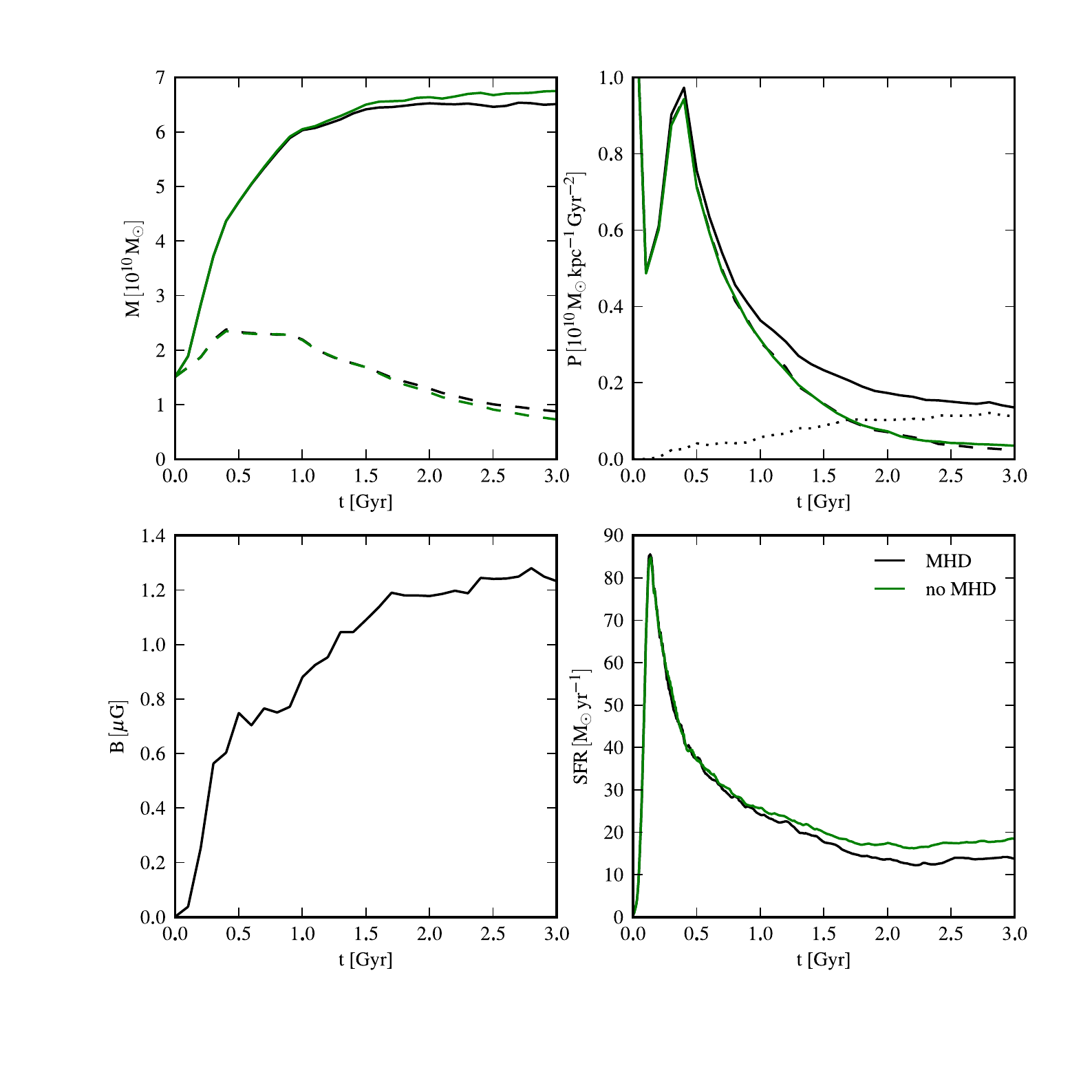}
  \caption{Time evolution of different quantities of a
    $10^{12}\,\mathrm{M_\odot}$ galaxy with (black lines) and without
    magnetic field (black lines) for a mass resolution of $2.1 \times
    10^5\,\mathrm{M_\odot}$. The top left panel shows the total
    baryonic mass in stars and gas (straight line) and the gas mass
    (dashed line) within a radius of $15\,\mathrm{kpc}$. The top right
    panel shows the thermal pressure (dashed line), the magnetic
    pressure (dotted line) and the total pressure (straight line). The
    pressure is calculated as volume-weighted average in a sphere of
    radius $15\,\mathrm{kpc}$ around the center of the galaxy. The
    bottom left panel contains the volume-weighted average root mean
    square of the absolute value of the magnetic field within a radius
    of $15\,\mathrm{kpc}$.  Finally, the bottom right panel shows the
    total star formation rate in the whole simulation.}
  \label{fig:timedepCmp}
\end{figure*}
 
\begin{figure*}
  \centering
  \includegraphics[width=18cm,trim=40 50 5 30,clip]{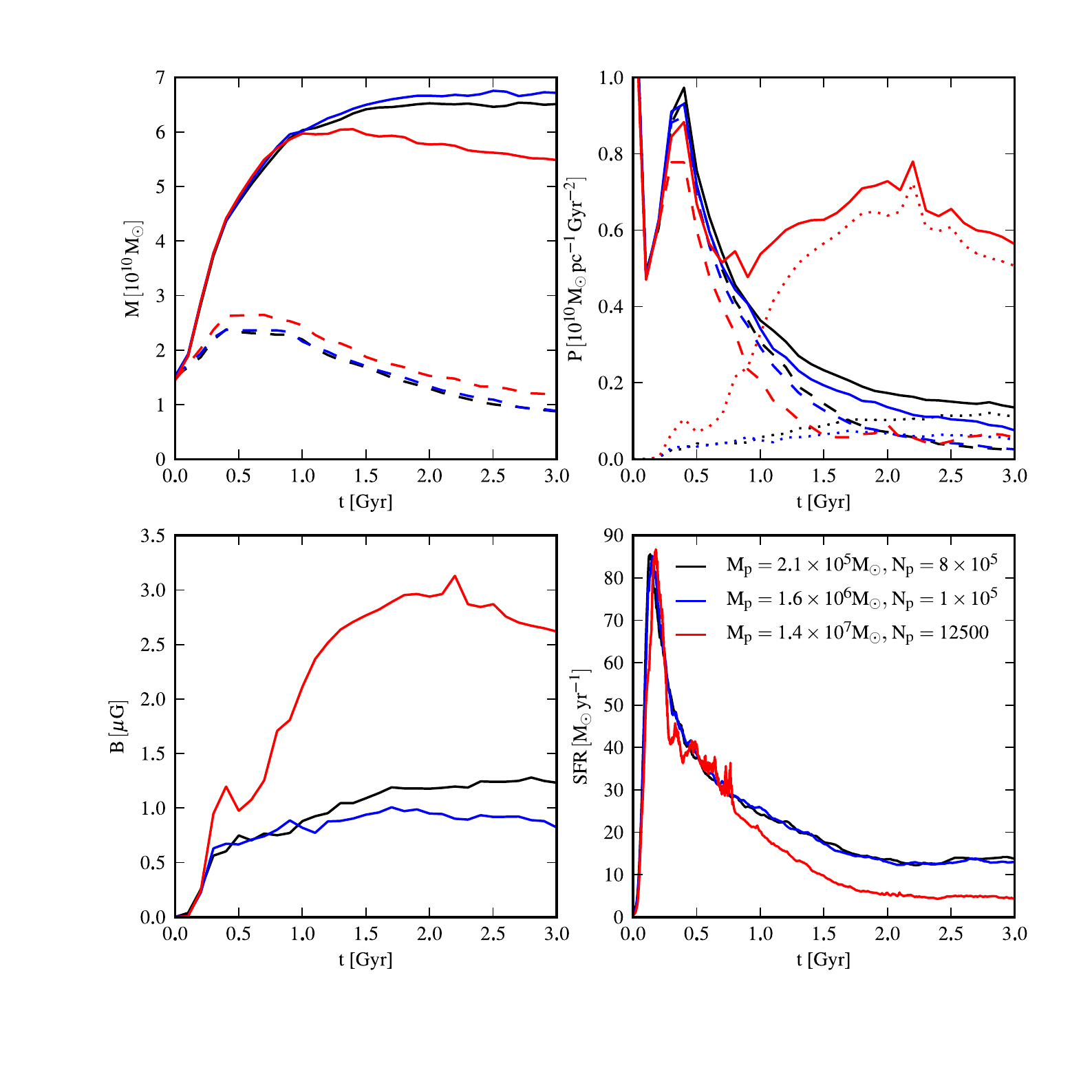}
  \caption{Resolution dependence of different quantities in
    simulations of a galaxy with a seed magnetic field of
    $10^{-9}\,\mathrm{G}$.  Shown are simulations with a mass
    resolution of $2.1 \times 10^5\,\mathrm{M_\odot}$ (black lines),
    $1.6 \times 10^6\,\mathrm{M_\odot}$ (blue lines), and $1.4 \times
    10^7\,\mathrm{M_\odot}$ (red lines).  The top left panel shows the
    total baryonic mass in stars and gas (straight line) and the gas
    mass (dashed line) within a radius of $15\,\mathrm{kpc}$. The top
    right panel shows the thermal pressure (dashed line), the magnetic
    pressure (dotted line) and the total pressure (straight line). The
    pressure is calculated as volume-weighted average in a sphere of
    radius $15\,\mathrm{kpc}$ around the center of the galaxy. The
    bottom left panel contains the volume-weighted average root mean
    square of the absolute value of the magnetic field within a radius
    of $15\,\mathrm{kpc}$.  Finally, the bottom right panel shows the
    total star formation rate in the whole simulation box.}
  \label{fig:timedepRes}
\end{figure*}
    
\begin{figure*}
  \centering
  \includegraphics[width=\textwidth]{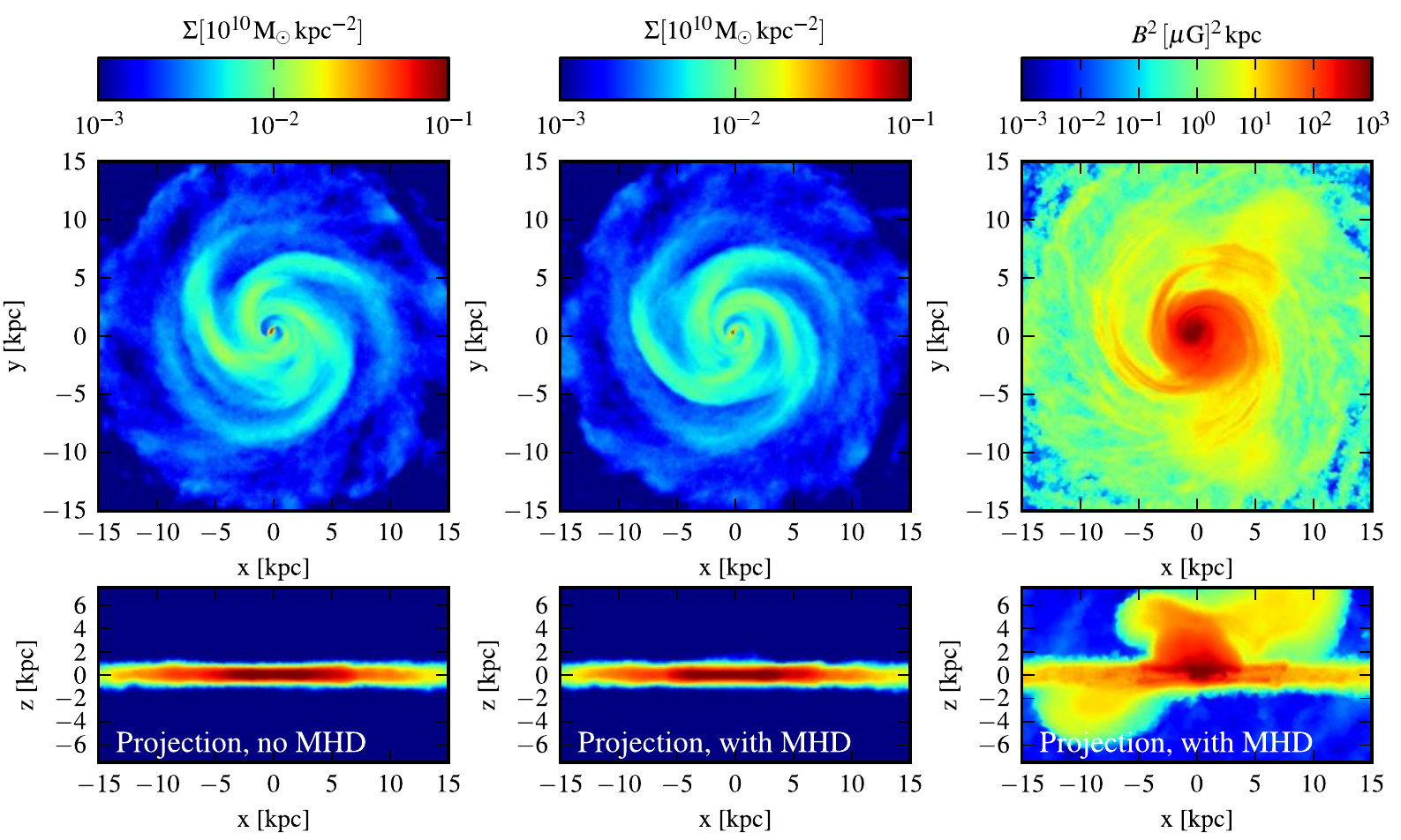}
  \includegraphics[width=\textwidth]{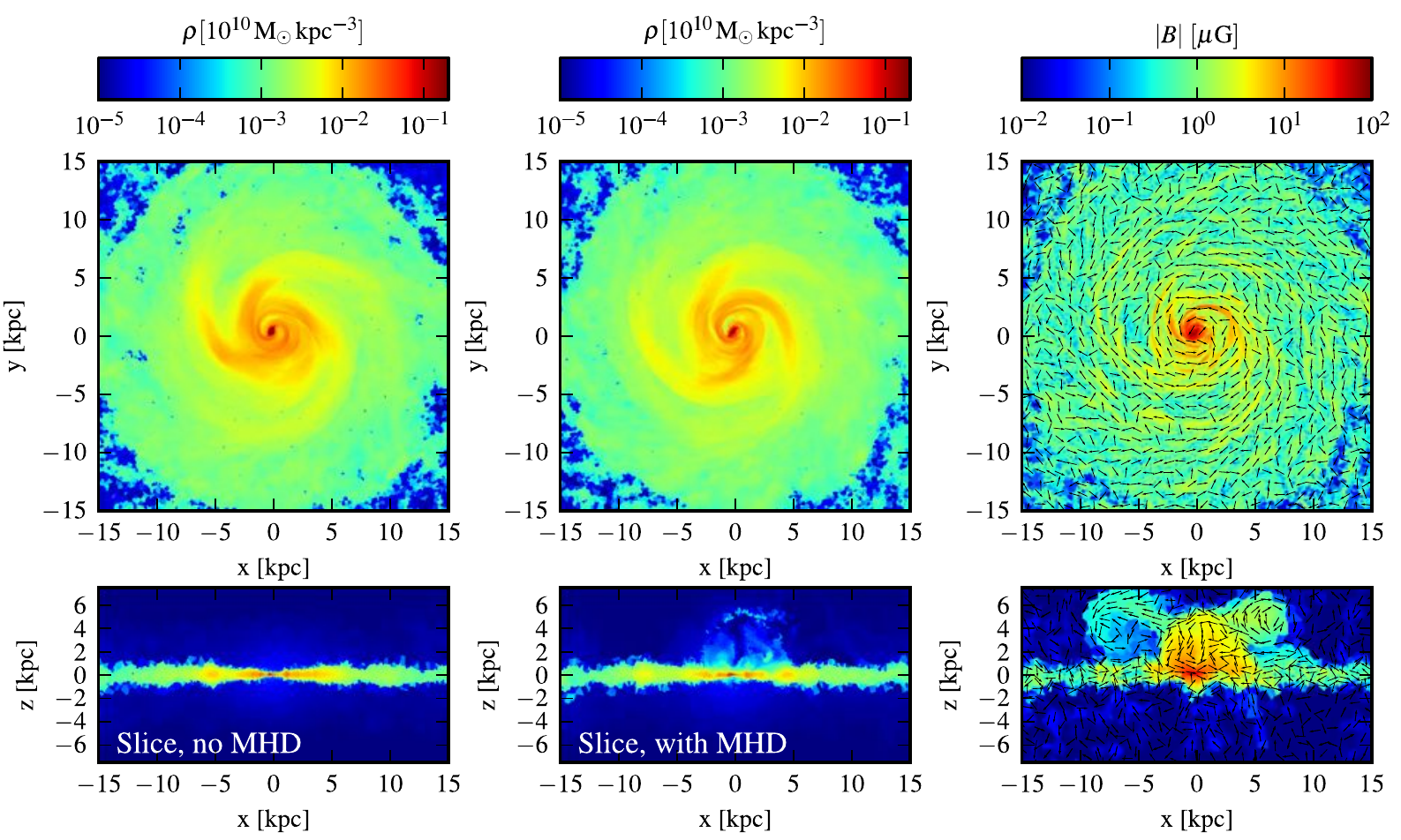}
  \caption{Gas density and magnetic field at $t=1\,\mathrm{Gyr}$. Top
    two rows show projections, bottom rows slices through the center
    of the galaxy. The left column shows the density for the
    simulation with a mass resolution of $2.1 \times
    10^5\,\mathrm{M_\odot}$ without a magnetic field. Center and left
    column show density and magnetic energy (for the projection) and
    magnetic field (on the slice) for the simulation with the same
    resolution but an initial magnetic seed field of
    $10^{-9}\,\mathrm{G}$.}
  \label{fig:gal10}
\end{figure*}
   
\begin{figure*}
  \centering
  \includegraphics[width=\textwidth]{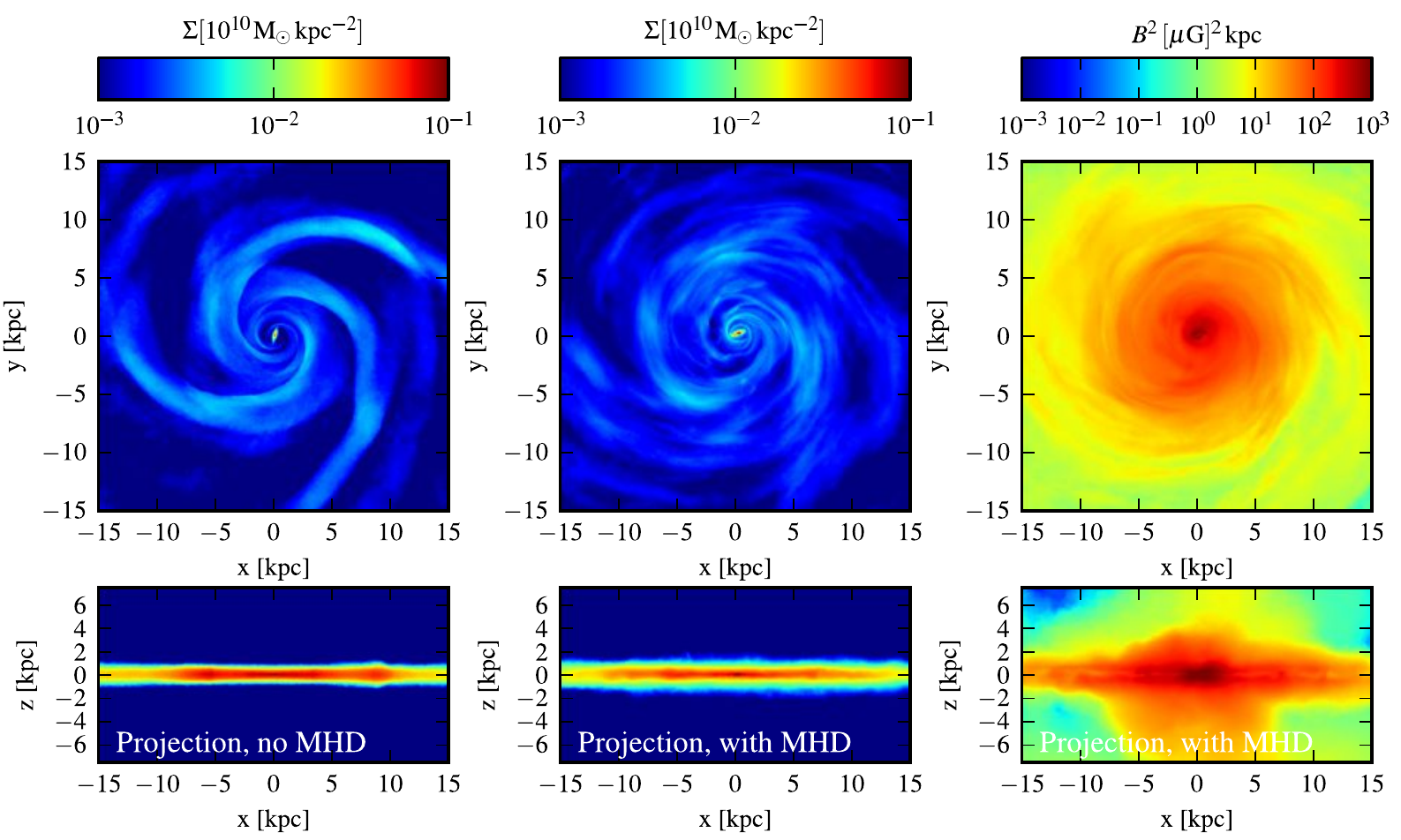}
  \includegraphics[width=\textwidth]{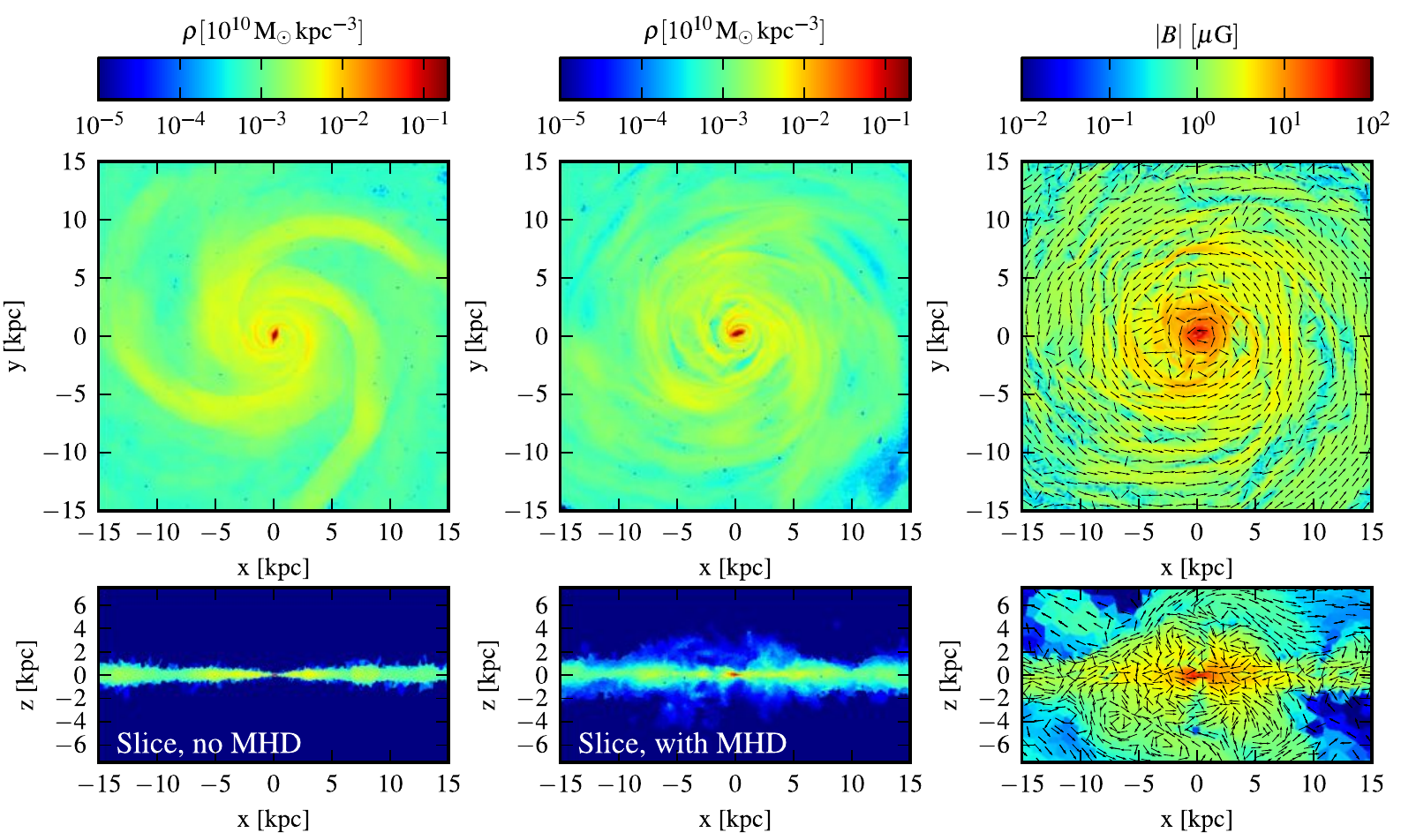}
  \caption{Gas density and magnetic field at $t=2\,\mathrm{Gyr}$. The
    two rows on top show projections, the bottom rows slices through
    the center of the galaxy. The left column depicts the density for
    the simulation with a mass resolution of $2.1 \times
    10^5\,\mathrm{M_\odot}$ without a magnetic field. Center and left
    column show density and magnetic energy (for the projection) and
    magnetic field (on the slice) for the simulation with the same
    resolution but an initial magnetic field of
    $10^{-9}\,\mathrm{G}$.}
  \label{fig:gal20}
\end{figure*}
    
\begin{figure*}
  \centering
  \includegraphics[width=\textwidth]{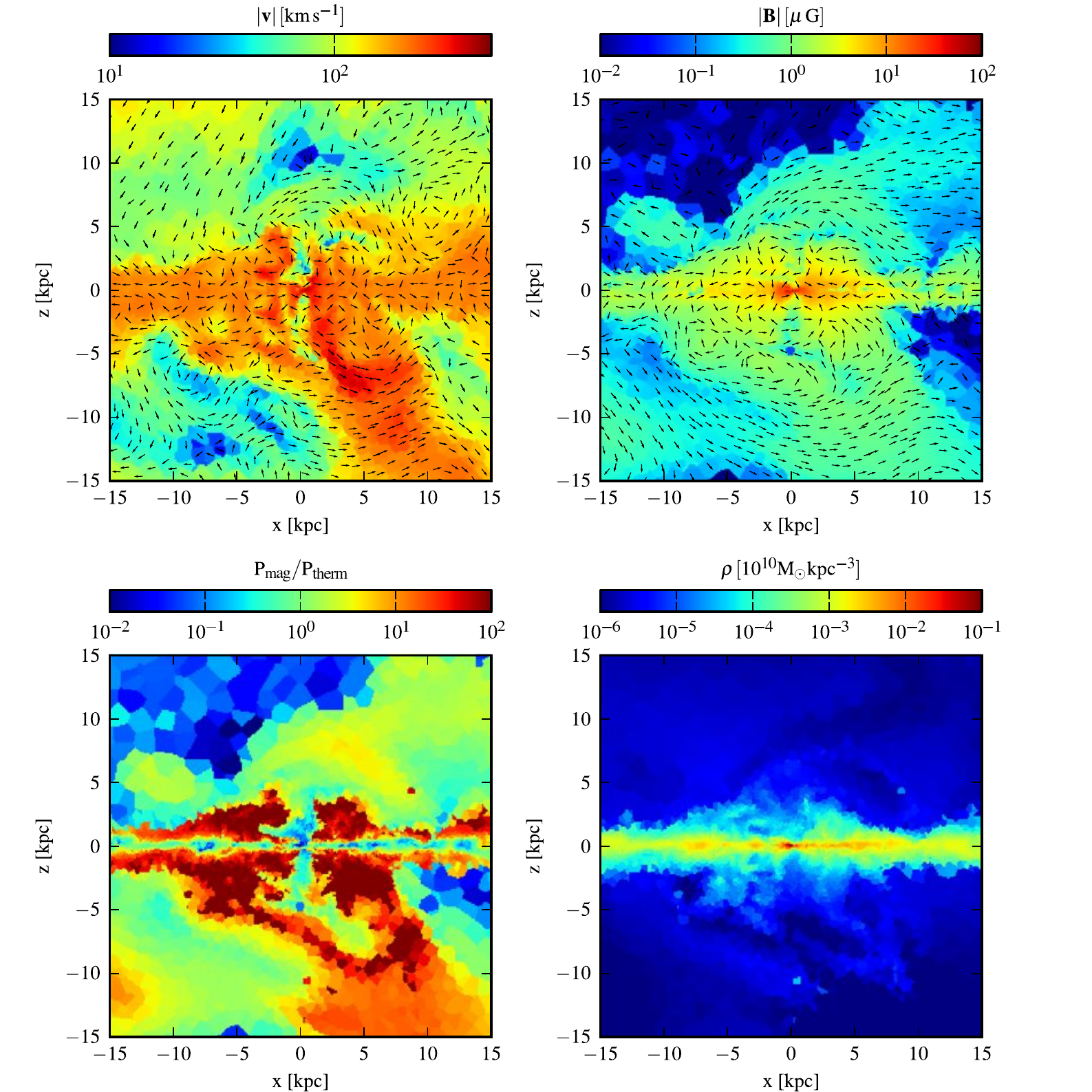}
  \caption{Properties of outflows above and below the disk at
    $t=2\,\mathrm{Gyr}$ for the simulation with a mass resolution of
    $2.1 \times 10^5\,\mathrm{M_\odot}$ with a magnetic field. Shown
    are the magnitude of the velocity (top left panel), the magnitude
    of the magnetic field (top right panel), the ratio of magnetic
    pressure to thermal pressure (bottom left panel) and the density
    (bottom right panel) in slices perpendicular to the disk.}
  \label{fig:atmo20}
\end{figure*}

\begin{figure}
  \centering
  \includegraphics[width=0.85\linewidth]{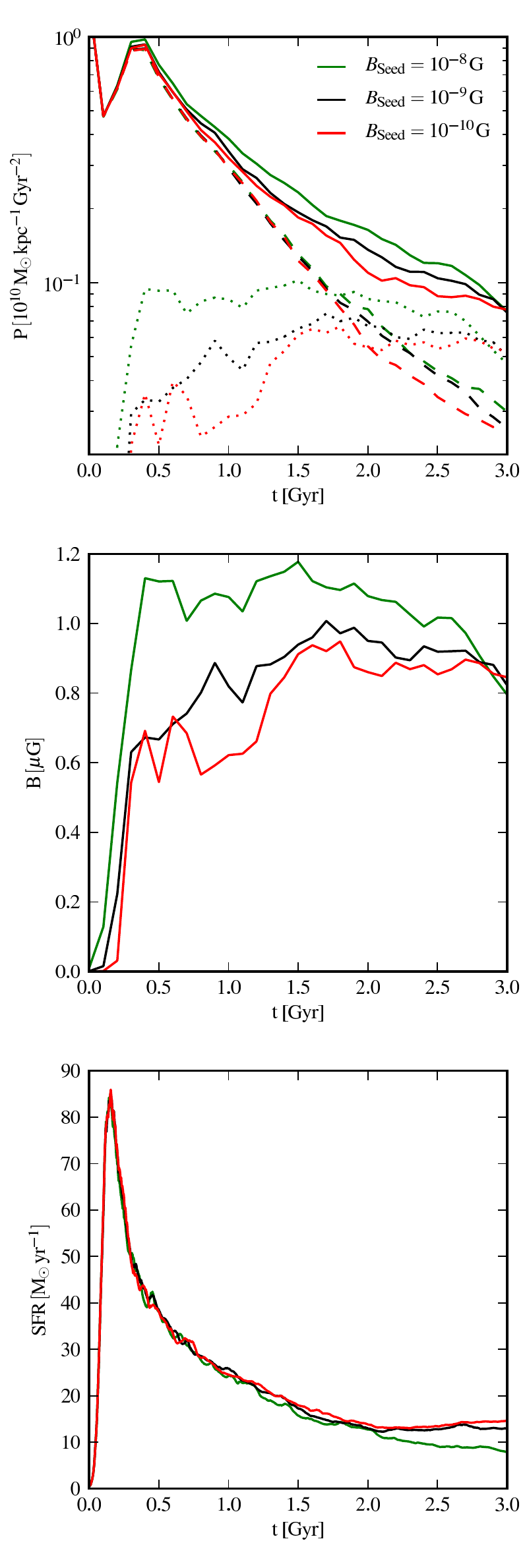}
  \caption{Time evolution of the pressure (top panel), the magnetic
  field (middle panel) and the star formation rate (bottom panel) for galaxies
  with seed magnetic fields from $10^{-8}\,\mathrm{G}$ to 
  $10^{-10}\,\mathrm{G}$. The straight, dashed and dotted lines in the
  top panel show volume-averaged total, thermal and magnetic pressure,
  respectively.}
  \label{fig:timedepB}
\end{figure}

\begin{figure}
  \centering
  \includegraphics[width=\linewidth]{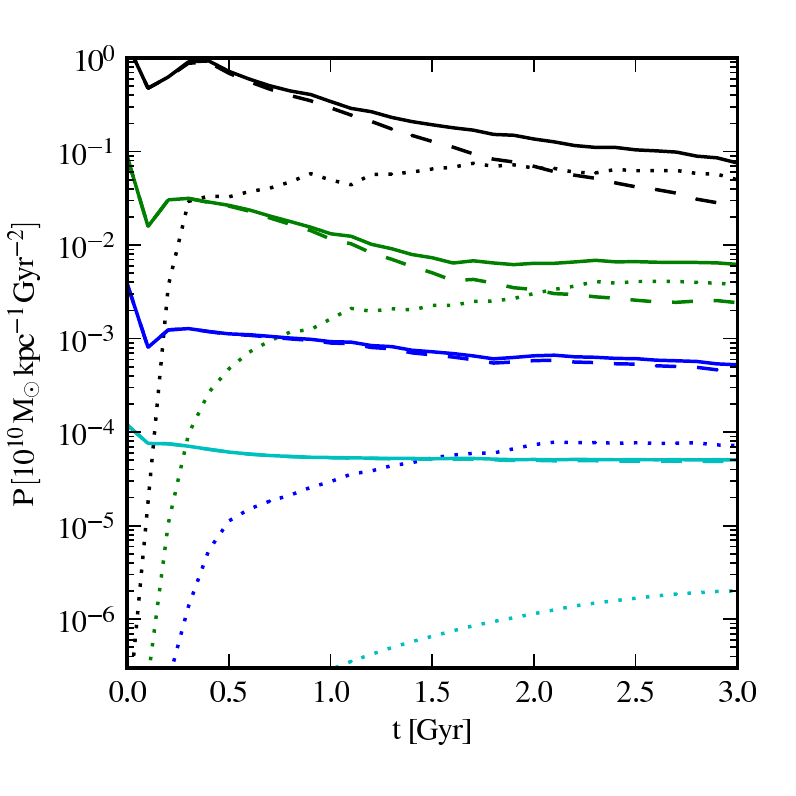}
  \caption{Time evolution of the volume-averaged total pressure
    (straight lines), thermal pressure (dashed lines) and 
    magnetic pressure (dotted lines). Black, green, blue and cyan
    lines represent galaxies with halo masses of
    $10^{12}\,\mathrm{M_\odot}$, $10^{11}\,\mathrm{M_\odot}$,
    $10^{10}\,\mathrm{M_\odot}$, and $10^{9}\,\mathrm{M_\odot}$,
    respectively.}
  \label{fig:timedepPresMass}
\end{figure}

\section{Simulation setup and additional physics} \label{sec:setup}

Our disk galaxy simulations start from slowly rotating spherical gas
clouds embedded in equilibrium into a collisionless dark matter halo
\citep[using the setup of][]{jubelgas2008a}. The dark matter halo is
modelled with a static background potential corresponding to a NFW
density profile \citep{navarro1997a} with a mass of $M_{200} =
10^{12}\,\mathrm{M_\odot}$ and a concentration parameter of $7.2$
roughly resembling the halo of the Milky Way.  We employ a spin
parameter of $\lambda = 0.05$ and a baryonic mass fraction of
$0.17$. In the simulations with magnetic fields, we typically
introduce a homogeneous seed magnetic field of $10^{-9}\,\mathrm{G}$
parallel to the $x$-axis. The same or similar seed field strengths
have also been used by \citet{wang2009b}, \citet{Kotarba2009} and
\citet{dubois2010a}. Indeed, intergalactic fields of this size are
expected in models of IGM magnetic seeding through outflows from dwarf
galaxies at high redshift \citep{Kronberg1999}.  Nevertheless, we have
also varied the strength of the seed field in a subset of our tests.
  
Gravitational forces are calculated with a standard tree-method
\citep[based on][]{Barnes1986, springel2005a} to account for the
self-gravity of the gas. The forces due to dark matter are determined
from the static background potential, neglecting a possible the
adiabatic contraction of the halo when baryons cool and settle in the
centre.  We use the \textsc{Arepo} code in its pseudo-Langrangian
configuration for our simulations. In this mode, the mesh-generating
points, which define the Voronoi cells, are moved with the local fluid
velocity, subject to small corrections to keep the shape of cells
reasonably regular. In addition, we use refinement and derefinement
operations where needed to ensure that the mass of the cells always
stays within a certain narrow range: if a cell contains more than
twice or less than half of the desired average mass per cell, we split
it into two cells or merge it with its neighbours, respectively. The
refinement and derefinement operations are carried out as described in
\citet{springel2010a} and \citet{Vogelsberger2012}.
   
The gas is allowed to cool radiatively, which eventually causes the
rotating gas sphere to develop a strong cooling flow and to form a
rotationally supported disk inside-out at the centre.  For simplicity,
we only include atomic cooling by helium and hydrogen, ignoring
molecules or metals. Therefore, the gas can only cool down to a
temperature of $10^4\,\mathrm{K}$.  We include star formation and
supernova feedback by means of a simple sub-resolution model
\citep{springel2003a}. This model assumes that star formation and
associated supernova feedback lead to a self-regulated multi-phase
interstellar medium in which cold molecular clouds are embedded in a
tenuous hot phase roughly at pressure equilibrium.  The pressurization
of the medium due to supernova feedback can can be described by an
effective equation of state. Stars are formed in a probabilistic
approach out of the medium, with a gas consumption time-scale
consistent with the observed \citet{Kennicutt1998} relation for
star-forming disk galaxies in the local Universe.  Once a star is
formed from a cell, we remove $90\%$ of the mass, momentum and
internal energy of the cell and create a collisionless star particle
at the position of the cell that inherits the removed mass and
momentum. Since the cell only retains $10\%$ of its mass, often a
derefinement of the cell is triggered in the next timestep, meaning
that the cell will be dissolved and its contents merged with its
neighbors.
   
We do not include any local supernova feedback in the form of
point-like energy depositions. Such local supernova feedback would
likely increase the amplification of the magnetic field, because of
the implied additional shearing motions and the higher turbulence in
the interstellar medium.
   
Also, we do not change the magnetic field of a cell when a star is
formed, similar to \citet{dubois2010a} in their AMR-code.  This has
the major advantage that the local structure of the magnetic field is
not affected by a star-forming event, and in particular, the
divergence of the magnetic field does not increase when a star is
formed. Note, however, that this is a conservative and non-trivial
assumption. For the mass resolution we can afford, each ``star
particle'' is a macro-particle that represents a whole stellar
population formed in a molecular cloud. In such a cloud, magnetic
fields can be strongly amplified by turbulence. At the same time, some
part of the magnetic field is locked up in stars, whereas another part
is pushed out of the cloud by ambipolar diffusion. Therefore, the
field dynamics in star-forming clouds is complicated and in reality it
may well be possible that the magnetic field surrounding a molecular
cloud is amplified during star formation. Similarly, but arguably less
likely, it is possible that a molecular cloud leaves a smaller
magnetic field behind after being depleted by star formation.
   
\section{Disk galaxy results}   \label{sec:results}
    
Without radiative cooling, the initial gas spheres in our simulations
remain in hydrostatic equilibrium. However, as we allow the gas
to cool, the pressure support in the centre is removed and the gas
sphere collapses. Since gas elements carry some angular momentum, they
settle into a dense, rotationally supported disk that can locally
fragment and form stars. The result is the formation of a disk galaxy
in a classic inside-out fashion \citep{Fall1980}. Similar numerical
experiments have frequently been used to study models of star
formation and feedback \citep[e.g.][]{springel2003a,
  jubelgas2008a}. Here, we are primarily interested in whether magnetic
fields can impact the dynamics of this disk formation scenario.
    
\subsection{Global evolution}

The global evolution of two galaxies with and without a magnetic seed
field is compared in Figure~\ref{fig:timedepCmp}. For the first
$1.5\,\mathrm{Gyrs}$ the evolution is quite similar. After only about
$200\,\mathrm{Myrs}$ the inner part of the gas sphere has collapsed
and formed a small disk in which a large starburst has ensued. In this
starburst, the star formation rate peaks just below
$100\,\mathrm{M_\odot\,yr^{-1}}$, but drops to about
$10-20\,\mathrm{M_\odot\,yr^{-1}}$ within a few hundred million years,
at which point a more or less constant, only slowly declining star
formation rate is reached.

The magnetic field is strongly amplified in the initial collapse of
the gas cloud and the first starburst, from an initial field strength
of $10^{-9}\,\mathrm{G}$ to an average field strength of the order of
$\mathrm{\mu G}$ in the inner parts, in agreement with observational
constraints \citep{beck2007a}. Much of the subsequent amplification
proceeds though an $\Omega$-type dynamo, made efficient through the
continued radial gas inflow combined with strong differential shear.
The magnetic field saturates at the level of few $\mathrm{\mu G}$, and
no significant further amplification takes place in our simulation in
places where the magnetic field has saturated. Once it has been
amplified to this saturation level, we find that it also contributes
significantly to the total pressure. Owing to the additional magnetic
pressure, it becomes slightly harder for the gas to form
stars. Therefore, as shown in Fig.~\ref{fig:timedepCmp}, the star
formation rate is lowered by about $30\%$ at late times in the
simulation with magnetic fields, compared to the run without magnetic
fields. At late times, after more than $1.5\,\mathrm{Gyrs}$, the
average magnetic pressure even begins to dominate the average thermal
gas pressure in parts of the star-forming gas disk.

\begin{figure*}
  \centering
  \includegraphics[width=18cm,trim=20 10 5 10,clip]{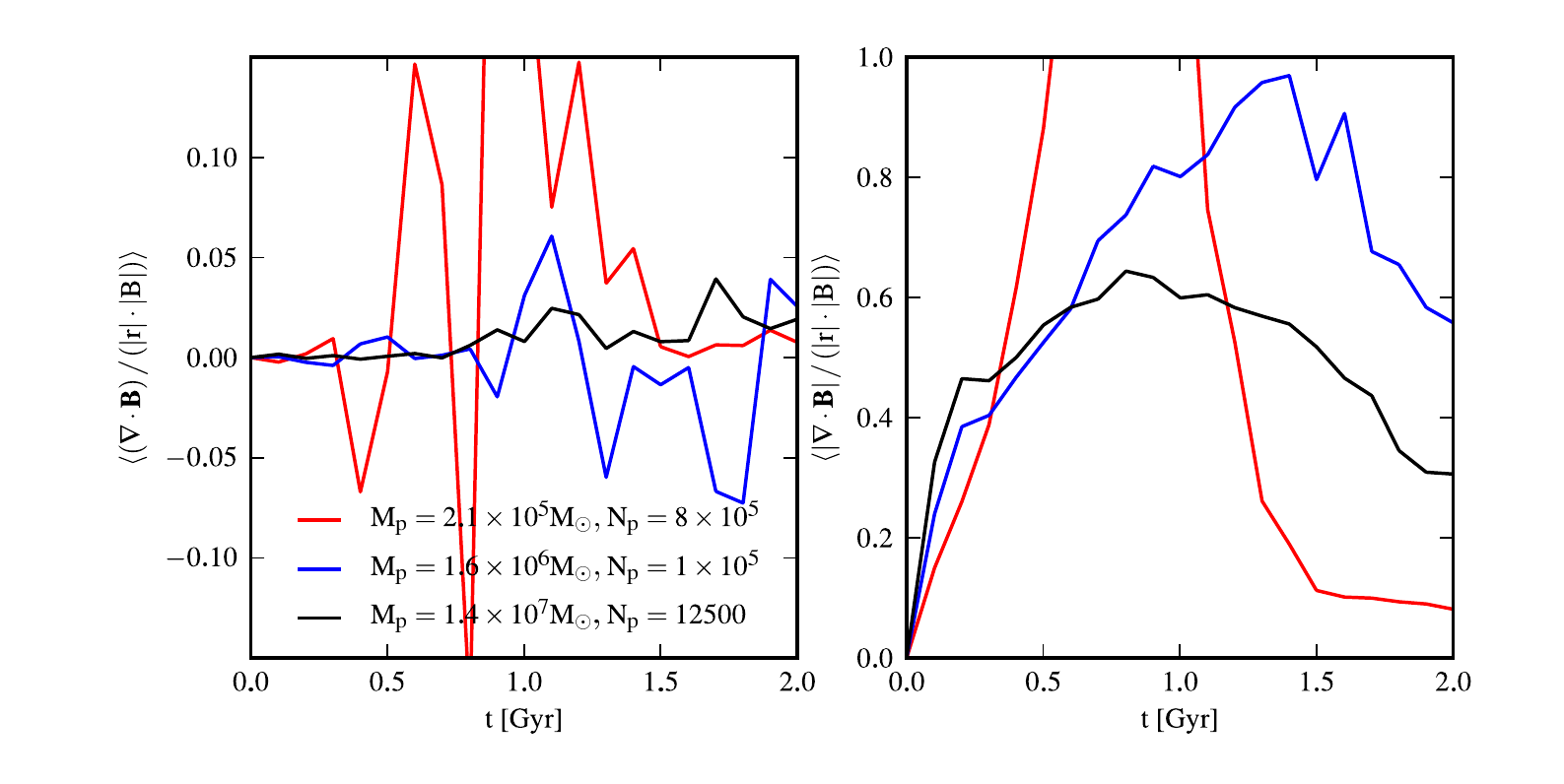}
  \caption{Time evolution of the relative divergence error of the
    magnetic field for the simulations with a mass resolution of $2.1
    \times 10^5\,\mathrm{M_\odot}$ (black lines), $1.6 \times
    10^6\,\mathrm{M_\odot}$ (blue lines), and $1.4 \times
    10^7\,\mathrm{M_\odot}$ (red lines). Left and right panels show
    the volume-weighted average divergence error and the
    volume-weighted average \textit{absolute} divergence error.}
  \label{fig:timedepErr}
\end{figure*}
   
\begin{figure*}
  \centering
  \includegraphics[width=18cm,trim=10 15 10 20,clip]{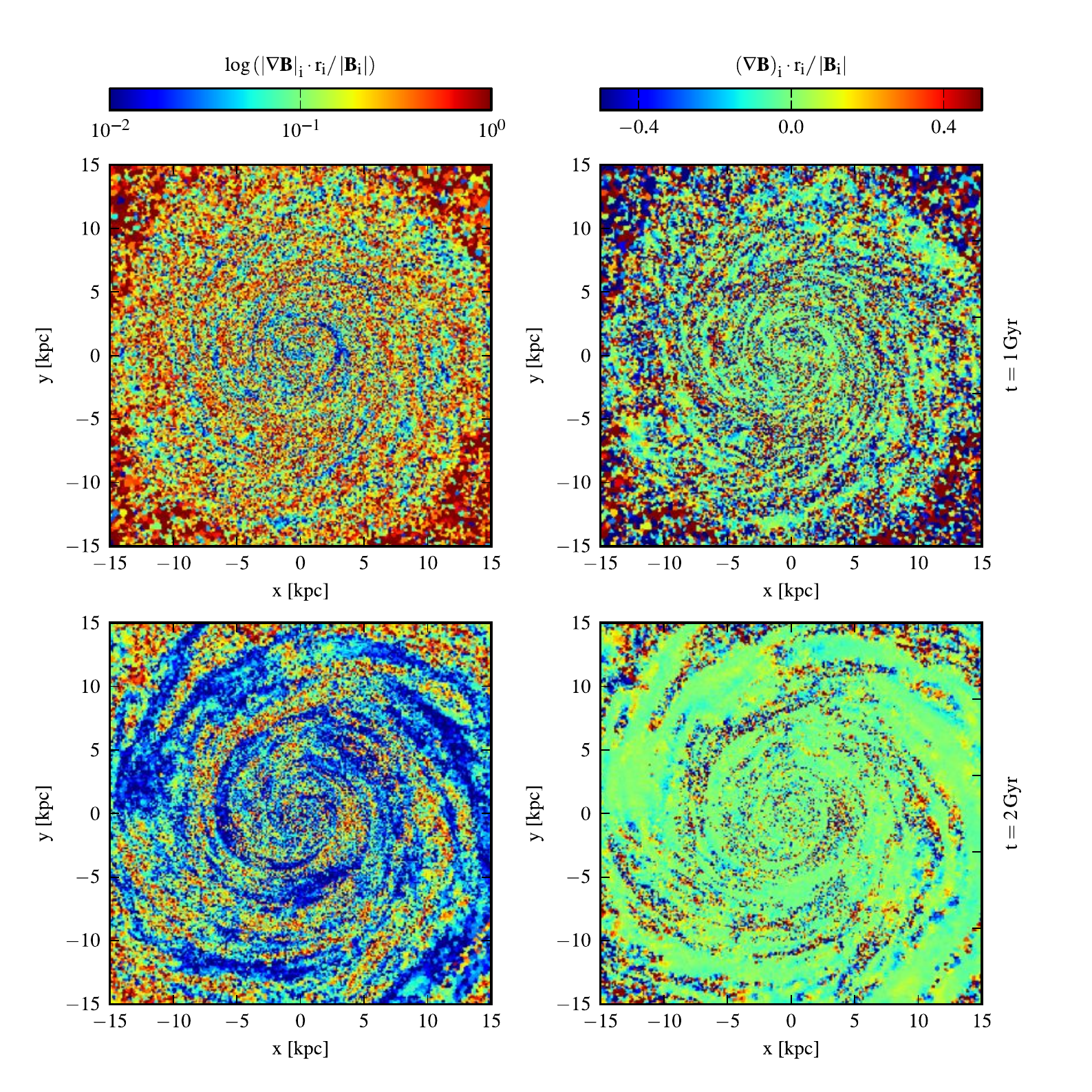}
  \caption{Relative divergence error of the magnetic field at
    $t=10\,\mathrm{Gyr}$ (top row) and at $t=20\,\mathrm{Gyr}$ (bottom
    row). The left column shows the absolute value of the relative
    divergence error, the right column shows the signed values.}
  \label{fig:errorgal}
\end{figure*}

\begin{figure*}
  \centering
  \includegraphics[width=18cm,trim=10 15 10 20,clip]{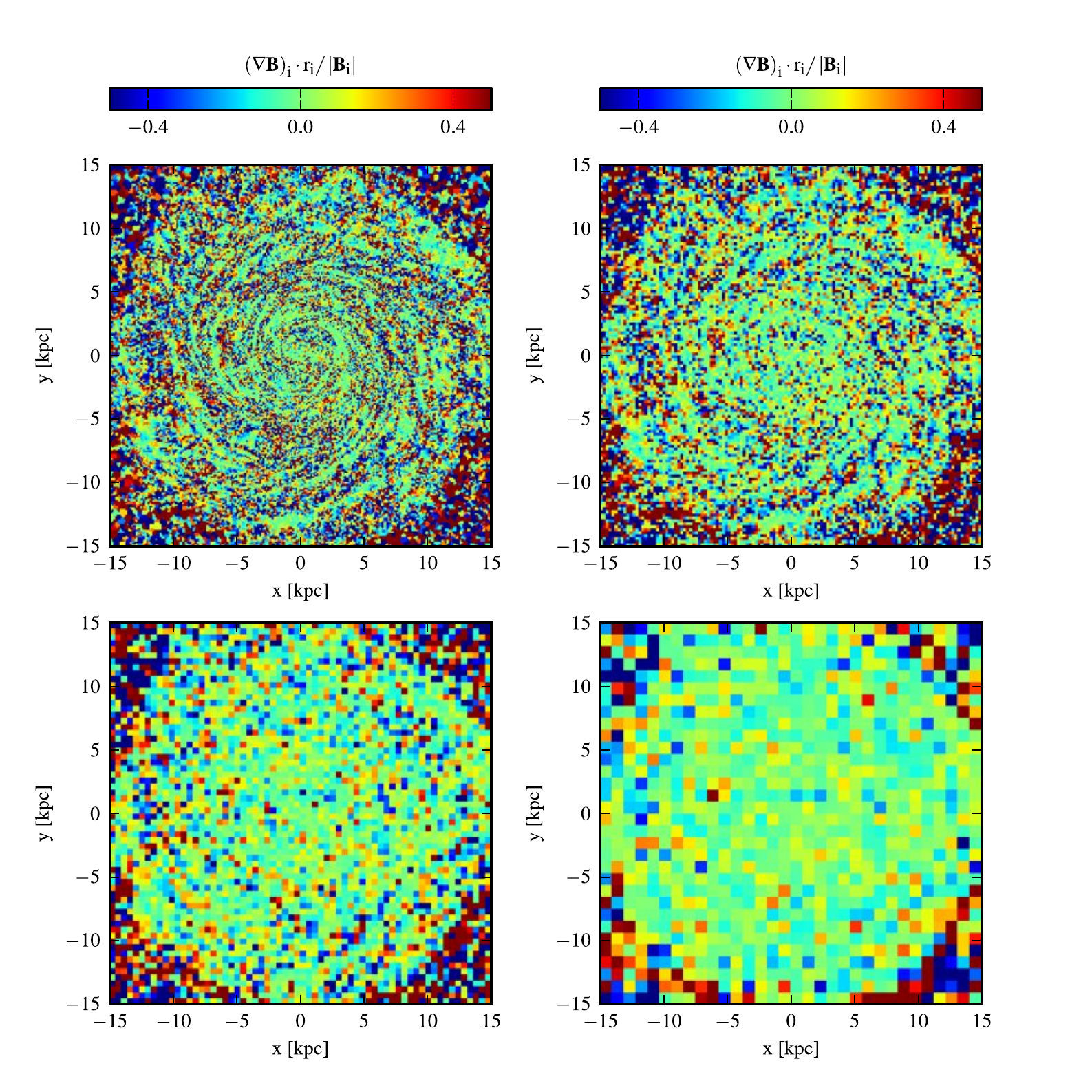}
  \caption{Relative divergence error of the magnetic field at
    $t=10\,\mathrm{Gyr}$. The top left panel shows the error sampled
    on a $256 \times 256$ grid. Top right, bottom left, and bottom
    right panel show the error from the top left panel averaged over
    $2\times 2$, $4 \times 4$, and $8 \times 8$ cells. This averaging
    rapidly reduces the residual local errors, indicating that the
    errors are locally compensated and do not induce incorrect
    large-scale magnetic dynamics.}
  \label{fig:errorav}
\end{figure*}

\subsection{Resolution dependence}

Before we proceed with further analysis, we investigate the
convergence of our results with numerical resolution. To this end, we
have run the simulation with magnetic field at three different
resolutions, using mass resolutions of $2.1 \times
10^5\,\mathrm{M_\odot}$, $1.6 \times 10^6\,\mathrm{M_\odot}$, and $1.4
\times 10^7\,\mathrm{M_\odot}$, respectively. The evolution of the
galaxy for these different resolutions is compared in
Figure~\ref{fig:timedepRes}.

The two higher resolution runs agree fairly well with each other. In
particular, the magnetic field saturates at about the same value and
it takes about the same time to reach saturation. In the lowest
resolution simulation, however, the magnetic field is amplified
somewhat faster, and saturates at an average value that is about a
factor of three larger compared to the higher resolution runs. The
magnetic pressure in this simulation also takes over the thermal
pressure earlier (at about $t=1\,\mathrm{Gyr}$) and the ratio of
average magnetic pressure over average thermal pressure is
significantly larger than unity. This finally leads to a considerably
stronger suppression of star formation, and a late-time star formation
rate of about one tenth of the late-time star formation rates in the
high-resolution runs.  

These differences suggest that in the low resolution run the
amplification of the magnetic field is artificially increased due to
poor resolution, caused in particular by relatively large divergence
errors (see Fig.~\ref{fig:timedepErr}).  We note however that the
impact of these errors on the amplification is very much smaller than
in the SPH calculations of \citet{Kotarba2009}, where order of
magnitude differences were found for different approaches to control
the divergence errors. In contrast to our results,
\citet{dubois2010a} have found slightly smaller amplification for
poor resolution, perhaps as a result of their accurate suppression of
divergence errors.  For isolated galaxies, the weak systematics we see
in our technique for low resolution is not a serious problem, because
such models can nowadays be simulated with at least the medium resolution we
use here. In cosmological simulations of structure formation where
galaxies form hierarchically, the reduced accuracy at low resolution
is potentially a more serious concern.  Here, the simulations 
always contain small progenitor galaxies which are initially resolved
only with typically very few cells. Accurately capturing the dynamics of these
systems when magnetic fields are involved may be quite challenging
with our techniques.

\subsection{Structure of the gas disk and its magnetic field} \label{sec:disk}

Projections of the gas density in the disk and maps of its magnetic
field are shown in Figures~\ref{fig:gal10} and \ref{fig:gal20}, after
one and two Gyrs, respectively.  After one Gyr, the inner part of the
gas halo has collapsed to a disk galaxy and converted a large fraction
of its gas into stars. The structure of the galaxy at this time is
very similar for the simulations with and without magnetic field. Even
the position and morphology of individual spiral arms are very
similar, and there is no difference in the thickness of the
disks. This is consistent with the result that the magnetic pressure
is on average still subdominant at this time, and therefore magnetic
fields are not expected to have a substantial effect on the dynamics
of the gas yet.
    
In the calculation with magnetic fields, the field has been amplified
to maximum values of up to $100\,\mathrm{\mu G}$ in the center of the
disk and to a few $\mathrm{\mu G}$ in its outer parts. The strength of
the magnetic field is correlated with the spiral arms, with a
significantly larger magnetic field in spiral arms than in between
arms \citep[consistent with][]{Kotarba2009}. Compared to the smooth
spirals arms seen in the density field, the magnetic field strength
shows a more patchy appearance with alternating regions of larger and
smaller magnetic field strength.  Interestingly, although we do not
include winds from supernovae, the simulation with magnetic fields
locally shows highly magnetized outflows from the disk which also
transport magnetic fields outwards. While overall the magnetic field
in the disk is aligned with the spiral structure of the disk, it also
shows a lot of field reversals, similar as in \citet{dubois2010a}. The
field strength is about a factor of 10 larger in the central part of
the disk compared to its outskirts. In the central part, where the
timescale on which differential rotation in the disk amplifies the
magnetic field is smaller, the magnetic field is also more regular. We
note that the structure of the simulated magnetic field is in good
agreement with observational data of spiral galaxies
\citep{beck2007a,Jansson2012}.
    
After an elapsed time of two Gyrs, the differences between the
galaxies simulated with and without magnetic fields have become
larger. By this time, the total gas mass in the disk has dropped by
about a factor of two, as a result of star formation and the reduced
supply of gas cooling out of the halo. The galaxy without magnetic
field still shows distinct spiral arms and also retains about the same
thickness it had one Gyr earlier.  In contrast, in the simulation with
a magnetic field, the individual spiral arms have mostly disappeared
to form a smoother disk with weaker residual spiral patterns. The
magnetic field in the disk is now very regular and well aligned with
the residual spiral structures of the disk. This change can be
explained in terms of the longer differential rotation time in the
outer parts of the disk.  Now after 2 Gyrs, differential rotation had
enough time to also amplify the magnetic field in the outer parts to
saturation values. Only in the very center of the disk a significant
radial component remains. Field reversals in the disk still exist but
have become much less frequent compared to the structure of the
magnetic field at time $t = 1\,{\rm Gyr}$.  In addition, the gas disk
is now noticeably thicker compared to the disk without magnetic
fields, as a result of the magnetic pressure contribution.

The magnetic field is primarily amplified by two processes in our
simulations. In the first few $100\,\mathrm{Myrs}$ (during the initial
starburst), adiabatic compression of the magnetic field dominates the
amplification. As material cools, it becomes denser until its density
reaches the star formation threshold. Then it forms stars and leaves a
highly amplified magnetic field behind. Therefore, the magnetic field
strength is initially correlated with the local star formation and is
highest at the center of the disk, where the compression and the star
formation rate is highest.  Later, after the total star formation in
the disk has dropped significantly, the amplification of the magnetic
field is dominated by shear motions in the differentially rotating
disk combined with radial inflows. This leads to a pronounced inside
out growth of the magnetic field in the disk, because the orbital
timescales increase for larger radii. The relentless shearing of the
magnetic field in the disk also make it ever more regular with time.
        
There are other processes which are likely important for the
amplification of magnetic fields in galaxies but which are not
included in our present simulations. Because we evolve an isolated
galaxy, it does not experience any large-scale shearing motions caused
by infalling material or mergers with other galaxies that could lead
to an additional amplification of the magnetic field. In addition, our
subresolution model for the interstellar medium does not directly
resolve interstellar turbulence, because we include supernova feedback
smoothly through an effective equation of state (see
Sec.~\ref{sec:setup}). Therefore, we miss a possible turbulent
amplification of the magnetic field on small scales.
    
In addition, we do not explicitly distinguish between neutral gas,
which should not feel the magnetic field directly, and ionized gas,
which interacts with the magnetic field. Instead, we neglect
resistivity effects and treat all gas in the limit of ideal
magnetohydrodynamics which will tend to overestimate the dynamical
effect of the magnetic pressure on the gas in the disk.
    
\subsection{Outflows} \label{sec:outflows}

The magnetized galaxy shows significant outflows of gas up to a few
kpc above the disk and below the disk, as shown in
Fig.~\ref{fig:atmo20}. These outflows are strongly magnetized and
their velocity vector is highly correlated with the magnetic field
lines. We note that unlike in the simulations of \citet{dubois2010a}
these outflows are not driven by supernovae. While supernova feedback
is implicitly included in our sub-resolution model for the ISM, it
does not create galactic winds; instead its effects are absorbed in the
applied effective equation of state.
    
The outflows are driven by low density, highly magnetized bubbles
which rise above the disk. In our simulation they reach typical
velocities of a few hundred $\mathrm{km\,s^{-1}}$, almost reaching
escape speed. The further they rise above the disk, the smaller their
velocity and magnetic field becomes until they finally fall back onto
the disk again. The outflows contain about $10^8\,\mathrm{M_\odot}$ of
gas at $t = 2\,\mathrm{Gyr}$.  Qualitatively, these outflows are
similar to galactic fountain models in which outflows are driven by
supernova-powered bubbles \citep[see,
e.g.,][]{shapiro1976a,oosterloo2007a,marinacci2011a}.  The quantitative
properties of the outflows, however, have to be taken with care, since
the resolution of our simulation is already rather low in these
regions.
    
\subsection{Varying the magnetic seed field}

To understand the effect of the magnetic seed field on the evolution
of the model galaxies, we repeat our simulation with the intermediate
resolution (which had a mass resolution of $1.6 \times
10^6\,\mathrm{M_\odot}$ and a seed field of $10^{-9}\,\mathrm{G}$)
with a larger and a smaller seed field of $10^{-8}\,\mathrm{G}$ and
$10^{-10}\,\mathrm{G}$, respectively.  The evolution of pressure,
magnetic field and star formation rate compared to the original
simulation is shown in Fig.~\ref{fig:timedepB}.  

Although the initial seed field differs by a factor of $10$ compared
with the original simulation, the evolution of the calculation with a
ten times smaller seed field is hardly changed. In this case, the
growth of the magnetic field is delayed by about $100\,{\rm Myrs}$
during the initial $300\, {\rm Myrs}$. After this time, when the
initial starburst subsides, the magnetic field in both simulations
converges to the same value and evolves basically identically,
indicating that the magnetic field saturated to a state independent of
the initial seed field. Because even directly after the starburst the
magnetic pressure is still too small compared with the thermal pressure
to be of dynamical relevance, the different choice of the seed field
in this case does not affect the evolution of the galaxy or the
evolution of the magnetic field once it is saturated.
    
The behavior of the simulation with a larger magnetic seed field
of $10^{-8}\,\mathrm{G}$, however, is slightly different. After the
initial starburst, the magnetic field is temporarily about twice as
large as in the original simulation (where a seed field of
$10^{-9}\,\mathrm{G}$ was used).  From this time onwards, it stays
roughly constant for two Gyrs at an elevated level until it finally
decreases to the same strength as in the other two simulations that
started with one or two order of magnitude weaker seed fields. Again,
since the magnetic pressure is much smaller than the thermal pressure
during the initial starburst and for most of the following time, this
difference has only a small effect on the total pressure or the star
formation rate. Only at late times, just before the magnetic field
decreases to the value found in the other simulations, the star
formation rate is suppressed slightly more than for the other two
calculations with smaller seed fields. We hence find that the
saturation strength of the magnetic field is almost completely
insensitive to the initial seed strength, apart from residual
differences stemming from the different times required for amplifying
the field from the initial strength to saturation.

\subsection{Halos of different masses}

Another important parameter in our simulations is the mass of the
simulated galaxy. To complement our set of simulations, we have run
three more simulations with $10^5$ gas cells, with the total mass of the
halo set to $10^{11}\,\mathrm{M_\odot}$,
$10^{10}\,\mathrm{M_\odot}$, and $10^{9}\,\mathrm{M_\odot}$,
respectively, and an initial magnetic field strength of
$10^{-9}\,\mathrm{G}$. The virial radii of the halos were scaled in a
self-similar fashion.

In Figure~\ref{fig:timedepPresMass}, we show the evolution of the
average pressure in the galaxies forming in these simulations and
compare it to our fiducial simulation with a halo mass of
$10^{12}\,\mathrm{M_\odot}$. The most important difference lies in the
relative strength of the magnetic pressure developing for galaxies of
different mass. For the two simulations with halo masses of
$10^{12}\,\mathrm{M_\odot}$ and $10^{11}\,\mathrm{M_\odot}$, the
magnetic pressures reaches equality with the thermal pressure after
about two Gyrs. However, for the two lower mass galaxies, the magnetic
field never becomes dynamically important. This difference from
self-similar behavior is to be expected because of the
non-selfsimilarity of cooling and star formation physics. Both the
cooling efficiency and the ability of a halo's gravitational potential
well to compress the gas of the interstellar medium against the
pressure delivered by supernovae vary strongly with halo mass.  In the
lower mass halos, the potentials become too shallow for compressing
the gas to very high densities, therefore the amplification of the
magnetic field by adiabatic compression of the seed field is
weaker. Furthermore, the much more anemic and thicker gas disks
developing in these halos reduce the amplification of the magnetic
field by differential shear. We thus expect that magnetic fields are
particularly important only for those halos that are also efficient
sites of star formation. According to recent abundance matching
constraints for the $\Lambda$CDM cosmology \citep{Guo2010,Moster2010},
these are primarily halos in the mass range $\sim
10^{11.5-12.5}\,\mathrm{M_\odot}$.

In the simulations with the $10^{12}\,\mathrm{M_\odot}$ and
$10^{11}\,\mathrm{M_\odot}$ galaxies, outflows are present as
discussed in Section~\ref{sec:outflows}, but they are absent in the
two low mass galaxies with halo masses of $10^{9}\,\mathrm{M_\odot}$
and $10^{10}\,\mathrm{M_\odot}$. This directly confirms that these
outflows only occur once the magnetic field becomes dynamically
important. We stress that these outflows are not caused by supernova
feedback, rather they can form when the magnetic field pressure
dominates over the thermal pressure of the gas in at least parts of
the disk.
    
\subsection{Divergence error} \label{sec:divbgal}

As discussed in Sec.~\ref{sec:divb}, our discretization scheme of the
MHD equations does not manifestly maintain a vanishing divergence of
the magnetic field to machine precision; rather a finite divergence
may appear, with the scheme preventing its further growth. Although
such small divergence errors do not cause any obvious artifacts even
in complex environments (including technically dicey steps such as
de-refinement and star formation), it is important to examine the
properties of the divergence error in more detail. We note that at the
very least it can alter the local jump conditions for the Riemann
solver, which in principle can sometimes lead to locally incorrect
fluxes, causing larger numerical noise or possibly more severe artifacts.
    
In Fig.~\ref{fig:timedepErr}, we show the evolution of the average
relative divergence error as a function of time for different
numerical resolutions in our default halo. With increasing resolution
the divergence error decreases, which indicates that it is dominated
by local noise in the magnetic field. This is also consistent with the
average divergence error being much smaller (by more than a factor of
10) than the average \textit{absolute} divergence error.
    
The spatial distribution of the divergence error is shown in
Figure~\ref{fig:errorgal}. Here it can be seen directly that the sign
of the divergence error is alternating locally on small scales. There
are basically no coherent patches of multiple cells with only positive
or negative divergence error. The magnitude of the divergence error
follows the pattern of the spiral arms, which can be easily understood
as a consequence of the divergence errors correlating with large
gradients of the magnetic field. In other words, large (unresolved)
gradients in the local magnetic field cause large divergence errors at
the same place. With higher resolution, the gradients are resolved
better and the divergence errors become smaller.
    
Another confirmation of the local nature of the divergence error can
be inferred from Figure~\ref{fig:errorav}, where local spatial
averages of the divergence error are shown. They quickly reduce the
error to the sub-percent level for progressively larger smoothing
regions. This suggests that non-zero errors in the field divergence do
not introduce changes in the large-scale dynamics of the magnetic
field.  Note also that the sizes of the divergence errors seen in our
simulations are smaller or at most of the same order than in
cosmological MHD simulations carried out with the latest generation of
SPH-MHD methods \citep{dolag2009a,beck2012a,stasyszyn2012a}, which
have similar properties regarding magnetic field divergence errors.

\section{Conclusions} \label{sec:summary}

In this study, we have examined the magnetic field amplification in
simple models of disk galaxy formation. Our goal has been to provide a
first exploration of the potential impact of magnetic fields for the
regulation of star formation in Milky Way-sized galaxies. We have
chosen isolated galaxies in order to allow a study of the magnetic
field evolution in a well-controlled setting, while still confronting
our numerical scheme with all the technical challenges (such as
dealing with star formation events and on-the-fly derefinements) it
has to cope with in future cosmological applications.

Another important aspect of our study lies in testing our new MHD
implementation in the moving-mesh code \textsc{Arepo}, which we
extended to cosmological simulations of galaxy formation that account
for radiative cooling and star formation. In contrast to our previous
method \citep{pakmor2011d}, we here use an 8-wave formulation
\citep{powell1999a} for dealing with the divergence constraint of the
magnetic field. This new implementation turns out to be much more
robust, whereas our previous MHD code could sometimes become unstable
in very dynamic environments unless very strict timestep contraints
were imposed.

We have shown in this paper that our new scheme produces competitive
results for test problems and is successful in keeping the divergence
error small. We have demonstrated that our new code is able to
simulate the magneto-rotational instability in disks without problems,
and that the correct linear growth rate is reproduced. Also, we have
shown that the new implementation prevents any noticeable artifacts in
simulations of disk formation.

For our simulations of isolated galaxy formation, we observe good
convergence for the strength of the magnetic field, the magnetic
pressure and the star formation rates, unless the resolution is very
low. In particular, we obtain converged results for our two high
resolution runs with mass resolutions of $2.1 \times
10^5\,\mathrm{M_\odot}$ and $1.6 \times 10^6\,\mathrm{M_\odot}$,
corresponding to an initial resolution of $8 \times 10^5$ gas cells
and $1 \times 10^5$ gas cells in the halo, respectively.  The magnetic
field saturates at a strength of about $10 - 100\,\mathrm{\mu\,G}$ in
the very center of the disk in a Milky Way-sized halo of mass
$10^{12}\,\mathrm{M_\odot}$, and at a few $\mathrm{\mu\,G}$ in the
main body of the disk. We find that the star formation rate and
structure of the gas disk is very similar for our runs with and
without magnetic fields until about $1\,\mathrm{Gyr}$. Later, the
magnetic pressure becomes comparable to the thermal pressure,
providing additional support for the gas and reducing the star
formation rate by about $30\%$.  It also causes changes the
structure of the disk, reducing the prominence of individual spiral
arms as the additional magnetic pressure produces slightly more
homogeneous disks. Interestingly, the magnetic field causes weak
outflows from the disks which rise several kpc above the disk before
their material falls back.

Simulations of halos with a reduced mass of
$10^{11}\,\mathrm{M_\odot}$ yield qualitatively very similar
results. Here the magnetic field is also amplified to a saturation
level, allowing it to affect the galaxy dynamics at late times. In
contrast, small halos with masses of $10^{10}\,\mathrm{M_\odot}$ and
$10^{9}\,\mathrm{M_\odot}$ did not amplify the magnetic field up to
equipartition with the thermal pressure, due to insufficient star
formation and gas compression. This can be understood as a consequence
of the shallow gravitational potential well and the modified cooling
efficiency in these systems.

Our results underline the potentially very important role magnetic
fields play in galaxy formation, especially in systems that have a
high efficiency of star formation, as here the gas is compressed
particularly strongly and the amplification through gas shearing
motions and inflows tends to be strong, too. In such galaxies,
magnetic fields of typical IGM strength are quickly amplified to
saturation values, such that the initial seed field strength is of
very limited importance and tends to be quickly forgotten.  It will be
very interesting to examine the influence of magnetic fields on
forming galaxies in cosmological hydrodynamic simulations of structure
formation, which we plan to study in future work. The present work has
demonstrated that such calculations are technically feasible and
sufficiently accurate with our present MHD implementation in
\textsc{Arepo}.
    
\section*{Acknowledgements}

R.P.~gratefully acknowledges financial support of the Klaus Tschira
Foundation. V.S.~acknowledges support by the DFG Research Centre
SFB-881 `The Milky Way System' through project A1.
    
  \bibliographystyle{mn2e}

  \label{lastpage}

\end{document}